\documentclass[11pt,a4paper]{article}
\usepackage{jheppub}
\usepackage{natbib}
\usepackage{graphicx}
\usepackage[dvipsnames]{xcolor}

\usepackage{mathtools}
\usepackage{hyperref}
\usepackage{float}
\usepackage{comment}
\usepackage{multirow}

\def\nat{Nature}

\def\mnras{MNRAS}

\def\aap{A\&A}

\definecolor{Blu}{rgb}{0.,0.,1.}

\newcommand{\CO}{\mathcal{O}}

\author[a,b]{Sunghyun Kang,}
\author[a,b]{Arpan Kar,}
\author[a,b]{Stefano Scopel,}
\affiliation[a]{Center for Quantum Spacetime, Sogang University, Seoul 121-742, South Korea}
\affiliation[b]{Department of Physics, Sogang University, Seoul 121-742, South Korea}
\emailAdd{francis735@naver.com}
\emailAdd{arpankarphys@gmail.com}
\emailAdd{scopel@sogang.ac.kr}

\title{Halo--independent bounds on the non--relativistic effective theory of WIMP--nucleon scattering from direct detection and neutrino observations}

\abstract{We combine experimental constraints from direct detection searches and from neutrino telescopes looking for WIMP annihilations in the Sun to derive halo--independent bounds on each of the 28 WIMP--proton and WIMP--neutron couplings of the effective non--relativistic Hamiltonian that drives the scattering process off nuclei of a WIMP of spin 1/2. The method assumes that the velocity distribution is normalized to one and homogeneous at the the solar system scale, as well as equilibrium between WIMP capture and annihilation in the Sun, and requires to fix the WIMP annihilation channels (we assume $b\bar{b}$). We consider a single non--vanishing coupling at a time, and find that for most of the couplings the degree of relaxation of the halo--independent bounds compared to those obtained by assuming the Standard Halo Model is with few exceptions relatively moderate in the low and high WIMP mass regimes, where it can be as small as a factor of $\simeq 2$, while in the intermediate mass range between 10 GeV and 200 GeV it can be as large as $\sim 10^3$. An exception to this general pattern, with more moderate values of the bound relaxation, is observed in the case of spin--dependent WIMP--proton couplings with no or a comparatively small momentum suppression, for which WIMP capture is strongly enhanced because it is driven by scattering events off $^1H$, which is the most abundant target in the Sun. Within this class of operators the relaxation is particularly small for interactions that are driven by only the velocity--dependent term, for which the solar capture signal is enhanced compared to the direct detection one, thanks to the highest speed of scattering WIMPs within the Sun due to the larger gravitational acceleration. } 

\begin{document}
\hspace*{87.0mm}{CQUeST-2022-0700}\\
\maketitle

\section{Introduction}
\label{sec:introduction}

A vast and global experimental effort has been undertaken in the last 30 years in the search of Weakly Interacting Massive Particles (WIMPs), the most popular candidates to provide the Cold Dark Matter that is supposed to have triggered galaxy formation and is believed to provide about 25\% of the density of the Universe under 
an invisible form~\cite{Planck:2018vyg}, 
only detected so far through its gravitational effects. In particular, a crucial physical process that is used to search for WIMPs is their scattering process off nuclear targets, that enters at the same time in Direct Detection (DD) experiments, that search for the recoil energy of nuclei in solid--state, 
liquid and gaseous detectors in underground laboratories shielded 
against cosmic rays~\cite{DD_Goodman1984, DD_LEWIN1996, JUNGMAN1996, DD_Schumann2019, Snowmass_Leane2022}, 
or in experiments searching for neutrinos produced by WIMP annihilation inside celestial bodies (Earth, Sun), 
where the WIMPs are accumulated after being captured through 
the same WIMP--nucleus scattering process 
that enters DD~\cite{cap_nu_sun_PhysRevLett1985, cap_nu_sun_HAGELIN1986, dm_cap_nu_SREDNICKI1987, Jungman:1994jr, idm_sun_Catena_2018}.

In both cases two classes of major uncertainties arise when it comes to compare the calculation of expected signals to the experimental data: the nature of the WIMP--nucleus interaction, and the WIMP speed distribution $f(u)$ in the reference frame of the Solar system that determines the WIMP incoming flux\footnote{Neglecting the relative velocity between the Earth and the Sun both present direct detection experiments and signals from WIMP capture in the Sun are sensitive to the speed distribution $f(u) \equiv \int d\Omega f(\vec{u}) u^2$.}. 
Indeed, for a long time experimental results of WIMP searches have been interpreted under specific assumptions about these two aspects. As far as the WIMP--nucleus interaction is concerned, the most common choices have been either a spin--independent (SI) WIMP--nucleus cross section, with a scattering amplitude proportional to the atomic mass number of the target, or  a spin--dependent (SD) interaction with a scattering amplitude proportional to the coupling between the spins of the WIMP and of the nucleus. Both types of interactions arise in popular extensions of the Standard model, such as Supersymmetry~\cite{JUNGMAN1996}. On the other hand, as for the $f(u)$, both early analytical estimations~\cite{violent_relaxation} and more recent numerical models of Galaxy formation~\cite{VDF_Lacroix2020, VDF_Lopes2020} are compatible to a 
Maxwellian in the galactic halo rest frame~\cite{SHM_1986, SHM_1988}, at least for speeds that are not far larger than a speed dispersion estimated to be of the order of $\simeq$ 300 km/s from the measurement of the galactic rotation curve , assuming hydrodynamic equilibrium between the pressure of the WIMP gas and the gravitational pull toward the center. This simple scenario, that predicts a flat rotation curve in agreement with observation, is also indicated as the Standard Halo Model (SHM).
As a consequence the default way to present the results of WIMP searches is still today based on providing upper bounds on the spin--independent and spin--dependent WIMP--nucleus cross section from direct detection and neutrino signal  under the assumption of a Maxwellian speed distribution $f_M(u)$. 

The non–-observation of new physics at the Large Hadron Collider (LHC) has increasingly constrained the most popular Dark Matter (DM) candidates predicted by extensions of the Standard Model, prompting the need to use bottom--up approaches that go beyond the SI/SD scenario.
Since the DD process is non--relativistic (NR), on general grounds the WIMP--nucleon interaction can be parameterized with an effective Hamiltonian ${\bf\mathcal{H}}$ that complies with
Galilean symmetry.  The effective Hamiltonian ${\bf\mathcal{H}}$ to zero--th order in the WIMP--nucleon relative velocity $\vec{v}$ and momentum transfer $\vec{q}$ consists of the usual spin--dependent (SD) and spin--independent (SI) terms. For WIMPs of spin 0 and 1/2 such Hamiltonian has been systematically extended to first order in the WIMP velocity $\vec{v}$ in Refs.~\cite{nreft_haxton1,nreft_haxton2}:

\begin{equation}
    {\cal H}=\sum_{\tau=0,1}\sum_{i=1}^{15} c_i^\tau {\cal O}_i,
    \label{eq:H}
\end{equation}
\noindent where the 14 Galilean--invariant operators ${\cal O}_i$ are listed in Table~\ref{tab:operators}.
In Eq.~(\ref{eq:H}) $c^{\tau}_i$ are the Wilson coefficients, with $\tau$ (= 0,1) 
the isospin, that can be arbitrary functions of the exchanged momentum $q$. 
For definiteness in our analysis we will consider constant Wilson coefficients, which correspond to a contact interaction, and elastic scattering. 
The coefficients $c^{\tau}_i$ can be converted into those for 
protons and neutrons through $c_i^p=c_i^0+c_i^1$ and $c_i^n=c_i^0-c_i^1$.

On the other hand, although the Isothermal Model provides a useful zero--order approximation to describe the WIMP speed distribution, numerical simulations of Galaxy formation can only shed light on statistical average properties of galactic halos, whilst our lack of information about the specific merger history of the Milky Way prevents us to rule out the possibility that the $f(u)$ has sizeable non--thermal components. Indeed, the growing number of observed dwarf galaxies hosted by the Milky Way suggests that our halo is 
not perfectly thermalized~\cite{Gaia_2018Nature, Gaia_2018MNRAS, Gaia_Myeong_2018, Gaia_Koppelman_2019, Gaia_Necib_2019, Gaia_Necib_2020, Gaia_OHare_2020}, 
and the more so should be expected in the high--speed tail of the $f(u)$ to which, for instance, DD signal are particular sensitive for light WIMP masses~\cite{DEAP2020, DD_Gaia_Bozorgnia2019}.
Based on the above considerations several attempts have been made to develop halo--independent approaches with the goal to remove the dependence of the experimental bounds on the choice of a specific speed distribution $f(u)$~\cite{halo_independent_2010, halo_independent_Fox_2010, halo_uncertainty_Frandsen2011, astrophysics_independent_Herrero-Garcia2012, halo_independent_DelNobile_2013, halo_independent_Fox2014, halo_independent_Feldstein2014, halo_independent_Scopel_inelastic_2014, halo_independent_Feldstein2014_2, halo_independent_Bozorgnia2014, halo_independent_Anderson2015, Halo-independent_Ferrer2015, halo_independent_Kahlhoefer, Gondolo_Scopel_2017, halo_independent_Catena_Ibarra_2018, velocity_uncertainty_Ibarra2018, velocity_independent_2019}.

Halo--independent techniques have been mainly developed in the context of direct detection. 
In one approach the factorization of a common generalized halo function~\cite{halo_independent_Fox_2010, halo_independent_DelNobile_2013, halo_independent_Scopel_inelastic_2014} allows to determine the scaling of expected signals in different detectors in a halo--independent way; an alternative strategy is to maximize the signal in one detector in compliance to a set of constraints from other detectors through the parameterization of the speed distribution in terms of a superposition of streams, $f(u)=\sum_{i=1}^{N} \delta(u-u_i)$. In particular, the use of linear algebra theorems~\cite{Gondolo_Scopel_2017, halo_independent_Kahlhoefer} allows to prove that in order to bracket the maximal variation of an expected signal the number $N$ of streams 
needs to be equal to the number of constraints. 
Both such methods have been applied to assess in a halo--independent way the compatibility of an experimental excess, such as the DAMA modulation effect~\cite{dama_libra_phase2}, with the constraints from other detectors~\cite{halo_independent_DelNobile_2013, halo_independent_Catena_Ibarra_2018}.  On the other hand, in absence of a clear excess an alternative strategy is to work out the most conservative bounds from null searches compatible with the only constraint:

\begin{equation}
    \int_{u=0}^{\infty} f(u) du = 1,
    \label{eq:f_normalization}
\end{equation}

\noindent but allowing for any possible speed profile of the distribution.

In the case of this latter approach WIMP direct searches run into a crucial limitation: all DD experiments are characterized by a recoil energy threshold $E_R^{th}$ that for a given WIMP mass converts into a speed threshold $u^{\rm DD}_{th}$ below which the sensitivity to the WIMP flux vanishes. As a consequence no conservative bounds can be established from existing DD experiments because the latter cannot probe the full range of WIMP speeds. In particular any functional form of the $f(u)$ for which $\int_{u=0}^{u^{\rm DD}_{th}} f(u) du = 1$, and, consequently, $\int_{u=u^{\rm DD}_{th}}^{\infty} f(u) du = 0$, corresponds to a vanishing expected signal in all existing DD experiments.

A possible solution to this problem is provided by combining the constraints from DD with those from the expected neutrino signal from WIMPs captured in the Sun~\cite{NT_DD_Kavanagh2014, NT_DD_Blennow2015}. 
Indeed, while capture in the Sun is suppressed at high WIMP incoming speeds, it is favoured for low (even vanishing) ones, because in the latter case it is easier for a slow WIMP to be scattered below the escape speed in order to remain gravitationally trapped in the celestial body. Such complementarity between DD and capture in the Sun
was exploited in Ref.~\cite{Halo-independent_Ferrer2015} to develop a particularly straightforward method that allows to obtain conservative constraints that are independent of the $f(u)$ and only require the assumption~(\ref{eq:f_normalization}). For convenience, in the following we will refer to such procedure as the ``single stream method", and to the ensuing constraints as ``single--stream halo--independent" bounds.

The applicability of the single--stream method of Ref.~\cite{Halo-independent_Ferrer2015} is limited to the case when both the DD and the neutrino signals are proportional to a single cross section or coupling. On the other hand the method cannot be used when the WIMP--nucleus scattering process is driven by the effective Hamiltonian of Eq.~(\ref{eq:H}), in presence of more that one effective operator ${\cal O}_i$.  
The ${\cal O}_i$ operators are nevertheless the most general building blocks of the low-energy limit of any ultraviolet theory, so that a discussion of the single stream method when the WIMP--nucleus interaction is driven by each of them is crucial for the interpretation of more general scenarios containing the sum of several non--relativistic operators.

As a consequence, in the present paper we wish to discuss the halo--independent single--stream bounds on each of the effective WIMP--proton and WIMP--neutron couplings $c_i^{p}$ and $c_i^{n}$, when such coupling is assumed to be the only non--vanishing one in Eq.~(\ref{eq:H}). In order to do so in our quantitative discussion we will combine DD bounds from 
XENON1T~\cite{xenon_2018}, PICO--60 ($C_3F_8$)~\cite{pico60_2019} and PICO--60 ($CF_3I$)~\cite{pico60_2015} to the bounds on WIMP capture in the Sun from neutrino telescopes (NTs) 
IceCube~\cite{IceCube:2016} and
Super-Kamiokande~\cite{SuperK_2015}. 
  
The plan of the paper is the following. In Section~\ref{sec:nreft} we briefly summarize the formalism of WIMP--nucleus scattering in WIMP--nucleon non--relativistic effective theory, and provide the corresponding expressions for the DD and WIMP capture signals in Sections~\ref{sec:dd} and \ref{sec:capture}.  In Section~\ref{sec:single_stream} we outline the single--stream method of Ref.~\cite{Halo-independent_Ferrer2015};  Section~\ref{sec:analysis} contains 
the results of our quantitative analysis. In particular 
the main results of the paper are shown in Figs.~\ref{fig:SI_coupling_mx} and \ref{fig:SD_coupling_mx}, where we systematically apply the procedure of Section~\ref{sec:single_stream} to calculate the conservative upper bound on each of the effective couplings $c_i^{p,n}$ of the Hamiltonian of Eq.~(\ref{eq:H}).
Our Conclusions are contained in Section~\ref{sec:conclusions}. Finally, in Appendix~\ref{app:experiments} we provide the details about the implementation of the experimental bounds. 

\section{Elastic WIMP--nucleus scattering in non--relativistic WIMP--nucleon effective theory}
\label{sec:nreft}

\begin{table}[]
\begin{center}
\begin{tabular}{|l|l|l|l|l|l|l|l|}
\hline\hline
\multicolumn{4}{l|}{\multirow{7}{*}{}} & \multicolumn{4}{l}{\multirow{7}{*}{}} \\
\multicolumn{4}{l|}{$ \CO_1 = 1_\chi 1_N$} & \multicolumn{4}{l}{$\CO_9 = i \vec{S}_\chi \cdot (\vec{S}_N \times {\vec{q} \over m_N})$} \\
\multicolumn{4}{l|}{$\CO_3 = i \vec{S}_N \cdot ({\vec{q} \over m_N} \times \vec{v}^\perp)$} & \multicolumn{4}{l}{$\CO_{10} = i \vec{S}_N \cdot {\vec{q} \over m_N}$} \\
\multicolumn{4}{l|}{$\CO_4 = \vec{S}_\chi \cdot \vec{S}_N$} & \multicolumn{4}{l}{$\CO_{11} = i \vec{S}_\chi \cdot {\vec{q} \over m_N}$} \\
\multicolumn{4}{l|}{$\CO_5 = i \vec{S}_\chi \cdot ({\vec{q} \over m_N} \times \vec{v}^\perp)$} & \multicolumn{4}{l}{$\CO_{12} = \vec{S}_\chi \cdot (\vec{S}_N \times \vec{v}^\perp)$} \\
\multicolumn{4}{l|}{$\CO_6=
  (\vec{S}_\chi \cdot {\vec{q} \over m_N}) (\vec{S}_N \cdot {\vec{q} \over m_N})$} & \multicolumn{4}{l}{$\CO_{13} =i (\vec{S}_\chi \cdot \vec{v}^\perp  ) (  \vec{S}_N \cdot {\vec{q} \over m_N})$} \\
\multicolumn{4}{l|}{$\CO_7 = \vec{S}_N \cdot \vec{v}^\perp$} & \multicolumn{4}{l}{$\CO_{14} = i ( \vec{S}_\chi \cdot {\vec{q} \over m_N})(  \vec{S}_N \cdot \vec{v}^\perp )$} \\
\multicolumn{4}{l|}{$\CO_8 = \vec{S}_\chi \cdot \vec{v}^\perp$} & \multicolumn{4}{l}{$\CO_{15} = - ( \vec{S}_\chi \cdot {\vec{q} \over m_N}) ((\vec{S}_N \times \vec{v}^\perp) \cdot {\vec{q} \over m_N})$} \\ \hline
\end{tabular}
\caption{Non--relativistic Galilean invariant operators for a WIMP of spin $1/2$ and up to linear terms in the WIMP velocity.}
\label{tab:operators}
\end{center}
\end{table}

When the scattering process is driven by the Hamiltonian of Eq.~(\ref{eq:H}) the WIMP--nucleus scattering amplitude takes the form~\cite{nreft_haxton1,nreft_haxton2} :

\begin{equation}
  \frac{1}{2 j_{\chi}+1} \frac{1}{2 j_{T}+1}|\mathcal{M}_T|^2=
  \frac{4\pi}{2 j_{T}+1} \sum_{\tau=0,1}\sum_{\tau^{\prime}=0,1}\sum_{k} R_k^{\tau\tau^{\prime}}\left [(c^{\tau}_i)^2, (v^{\perp})^2,\frac{q^2}{m_N^2}\right ] W_{T k}^{\tau\tau^{\prime}}(q).
\label{eq:squared_amplitude}
\end{equation}

\noindent In the expression above $j_{\chi}$ and $j_{T}$ are the WIMP
and the target nucleus spins, respectively, $m_N$ is the mass of nucleon and $q=|\vec{q}|$ is the magnitude of the transferred momentum, 
while the $R_k^{\tau\tau^{\prime}}$'s are WIMP response functions that can be found for instance in Ref.~\cite{nreft_haxton2} and that depend on the couplings $c^{\tau}_i$ as well as the transferred momentum $\vec{q}$ and on:

\begin{eqnarray}
(v^{\perp})^2 = v^2-v_{\rm min}^2 
\end{eqnarray}

\noindent where $v$ is the incoming WIMP speed, and:

\begin{equation}
v_{\rm min}^2=\frac{q^2}{4 \mu_{\chi T}^2}=\frac{m_T E_R}{2 \mu_{\chi T}^2},
\label{eq:vmin}
\end{equation}
\noindent (with $m_T$ and $\mu_{\chi T}$ the target nucleus mass and the WIMP--nuclear reduced mass) 
represents the minimal incoming WIMP speed required to
impart the nuclear recoil energy $E_R$. The WIMP response functions $R_k^{\tau\tau^{\prime}}$
can be decomposed in a velocity--independent and a velocity--dependent part:

\begin{equation}
R_k^{\tau\tau^{\prime}}=R_{0k}^{\tau\tau^{\prime}}+R_{1k}^{\tau\tau^{\prime}} (v^2-v_{\rm min}^2).
\label{eq:r_decomposition}
\end{equation}

Moreover, in equation
(\ref{eq:squared_amplitude}) the $W^{\tau\tau^{\prime}}_{T k}(y)$'s
are nuclear response functions and the index $k$ represents different
effective nuclear operators, which, under the assumption
that the nuclear ground state is an approximate eigenstate of $P$ and
$CP$, can be at most eight: following the notation in
\cite{nreft_haxton1,nreft_haxton2}, $k$ = $M$, $\Phi^{\prime\prime}$,
$\Phi^{\prime\prime}M$, $\tilde{\Phi}^{\prime}$,
$\Sigma^{\prime\prime}$, $\Sigma^{\prime}$,
$\Delta$, $\Delta\Sigma^{\prime}$. 
The $W^{\tau\tau^{\prime}}_{T k}(y)$'s are function of $y\equiv (qb/2)^2$, where $b$ is a parameter that depends on the size of the nucleus. For the target nuclei $T$ used in most direct detection experiments the functions $W^{\tau\tau^{\prime}}_{T k}(y)$, calculated using nuclear shell models, have been provided in
Refs.~\cite{nreft_haxton2,Catena_nuclear_form_factors}.

For each non--relativistic operator ${\cal O}_i$ the correspondence between  the WIMP response functions $R_{0k}^{\tau\tau^{\prime}}$, $R_{1k}^{\tau\tau^{\prime}}$ and the nuclear response functions $W^{\tau\tau^{\prime}}_{T k}$ is provided in Table~\ref{table:eft_summary}.

The common quantity required for the calculation of both the direct detection signal and that from WIMP capture in the Sun is then the differential cross section~\cite{nreft_haxton1,nreft_haxton2}:

\begin{equation}
\frac{d\sigma_T}{d E_R}=\frac{2 m_T}{4\pi v^2}\left [ \frac{1}{2 j_{\chi}+1} \frac{1}{2 j_{T}+1}|\mathcal{M}_T|^2 \right ].
\label{eq:dsigma_de}
\end{equation}

\begin{table}[t]
\begin{center}
{\begin{tabular}{@{}|c|c|c|c|c|c|@{}}
\hline
operator  &  $R^{\tau \tau^{\prime}}_{0k}$  & $R^{\tau \tau^{\prime}}_{1k}$ & operator &  $R^{\tau \tau^{\prime}}_{0k}$  & $R^{\tau \tau^{\prime}}_{1k}$ \\
\hline
$1$  &   $M(q^0)$ & - & $3$  &   $\Phi^{\prime\prime}(q^4)$  & $\Sigma^{\prime}(q^2)$\\
$4$  & $\Sigma^{\prime\prime}(q^0)$,$\Sigma^{\prime}(q^0)$   & - & $5$  &   $\Delta(q^4)$  & $M(q^2)$\\
$6$  & $\Sigma^{\prime\prime}(q^4)$ & - & $7$  &  -  & $\Sigma^{\prime}(q^0)$\\
$8$  & $\Delta(q^2)$ & $M(q^0)$ & $9$  &  $\Sigma^{\prime}(q^2)$  & - \\
$10$  & $\Sigma^{\prime\prime}(q^2)$ & - & $11$  &  $M(q^2)$  & - \\
$12$  & $\Phi^{\prime\prime}(q^2)$,$\tilde{\Phi}^{\prime}(q^2)$ & $\Sigma^{\prime\prime}(q^0)$,$\Sigma^{\prime}(q^0)$ & $13$  & $\tilde{\Phi}^{\prime}(q^4)$  & $\Sigma^{\prime\prime}(q^2)$ \\
$14$  & - & $\Sigma^{\prime}(q^2)$ & $15$  & $\Phi^{\prime\prime}(q^6)$  & $\Sigma^{\prime}(q^4)$ \\
\hline
\end{tabular}}
\caption{Nuclear response functions corresponding to each of the operators 
(listed in Table~\ref{tab:operators}), 
for the velocity--independent and the velocity--dependent components parts of the WIMP response function, decomposed as in Eq.(\ref{eq:r_decomposition}).
 In parenthesis the power of $q$ in the WIMP response function is shown.
  \label{table:eft_summary}}
\end{center}
\end{table}

\subsection{Direct detection}
\label{sec:dd}

In a direct detection experiment, the number of expected nuclear recoil events within visible energy, $E^{\prime}_1 \le E^{\prime} \le E^{\prime}_2$, is given by:

\begin{eqnarray}
R_{[E_1^{\prime},E_2^{\prime}]}&=&M{\tau_{\rm exp}}\int_{E_1^{\prime}}^{E_2^{\prime}}\frac{dR}{d
  E^{\prime}}\, dE^{\prime}, \label{eq:start}\\
 \frac{dR}{d E^{\prime}}&=&\sum_T\left(\frac{dR}{d E^{\prime}}\right )_T=\sum_T \int_0^{\infty} \frac{dR_{\chi T}}{dE_{ee}}{\cal
   G}_T(E^{\prime},E_{ee})\epsilon(E^{\prime})\label{eq:start2}\,d E_{ee}, \label{eq:diff_rate_eprime}\\
E_{ee}&=&Q(E_R) E_R \label{eq:start3}.
\end{eqnarray}

\noindent In the equations above, $M$ is the fiducial mass and $\tau_{\rm exp}$ the live--time of data taking while ${\cal G}(E^{\prime},E_{ee})$ represents energy resolution, $\epsilon(E^{\prime})$ the efficiency and $Q(E_R)$ the quenching factor, while:

\begin{equation}
\frac{d R_{\chi T}}{d E_R}=\frac{\rho_\odot}{m_\chi}\int_{v \geq v_{\rm min}(E_R)} dv f(v) v \sum_T N_T \frac{d\sigma_T}{d E_R} ,
\label{eq:dr_der}
\end{equation}

\noindent with $N_T$ the number of targets per unit mass in the detector. Here $\rho_\odot$ is the local density of DM for which we use the standard value 
$\rho_\odot = 0.3$ $\rm GeV/cm^{3}$~\cite{DD_LEWIN1996}. 
The function $f(v)$ is the normalised WIMP speed distribution in the reference frame of the Earth. 
Since we are only going to consider experiments that are sensitive to the time--average of the rate, 
in the equation above 
we identify the Earth reference frame to that of the solar system, effectively averaging away the effect of the Earth rotation around the Sun. Indicating with $u$ the WIMP speed in the reference frame of the Sun this is equivalent to make the identification $v \rightarrow u$, $v_{\rm min}\rightarrow u_{\rm min}$\footnote{When $E_R$ = $E_R^{th}$ is the experimental nuclear recoil threshold of a direct detection experiment the expression in Eq.~(\ref{eq:vmin}) represents $(v_{th}^{DD})^2$ = $(u_{th}^{DD})^2$.}. Then the total number of expected events in a DD experiment can be written as:

\begin{eqnarray}
R_{\rm DD} &=& M{\tau_{\rm exp}} \hspace{0.5mm} \left(\frac{\rho_\odot}{m_{\chi}}\right) 
\sum_{T} N_{T}  \int du \hspace{0.5mm} f(u) \hspace{0.5mm} u 
\hspace{0.5mm} 
\int^{E_R^{\rm max}}_{0} dE_R \hspace{0.6mm} \zeta_T(E_R,E_1^{\prime},E_2^{\prime}) \hspace{0.5mm} \frac{d\sigma_T}{dE_R},
\label{eq:DD_event}
\end{eqnarray}

\noindent where $E_R^{\rm max} = 2 \mu^2_{\chi T} u^2 / m_T$ and $\zeta_T$ indicates the response of detector that depends on the visible energy range, the energy resolution and the efficiency:

\begin{equation}
    \zeta_T=\int_{E_1^{\prime}}^{E_2^{\prime}} dE^{\prime}{\cal
   G}_T\left [E^{\prime},Q(E_R)E_R)\right]\epsilon(E^{\prime}) \rightarrow \epsilon(E_R)\Theta(E_{R,2}-E_R)\Theta(E_R-E_{R,1}).
    \label{eq:zt}
    \end{equation}
    
\noindent The three experimental collaborations that we will consider in Section~\ref{sec:analysis}  provide their results directly in terms of the true recoil energy $E_R$ and encode the energy range in the acceptance $\epsilon$ (see Appendix~\ref{app:experiments}). In this case $Q(E_R)$ =1, ${\cal G}_T=\delta ( E^{\prime}-E_R)$ and $E^{\prime}_{1,2}$ = $E_{R,1,2}$, and the response of the detector simplifies to the acceptance times a window function selecting the experimental energy bin, as shown in the last step of Eq.~(\ref{eq:zt}).  

\subsection{Capture in the Sun}
\label{sec:capture}

In the following we will assume that the WIMP capture and annihilation rates $C_\odot$ and $\Gamma_\odot$ are in equilibrium, i.e. $\Gamma_\odot$ = $C_\odot / 2$ (see Appendix~\ref{exp:NT}). In this case the neutrino flux from WIMP annihilations in the Sun is completely determined  by the capture rate given by~\cite{Gould:1987}: 

\begin{eqnarray}
C_\odot &=& \left(\frac{\rho_\odot}{m_{\chi}}\right) \hspace{0.5mm} \int du \hspace{0.5mm} f(u) 
\hspace{0.5mm} \frac{1}{u} \int^{R_\odot}_0 dr \hspace{0.5mm} 4 \pi r^2 \hspace{0.5mm} w^2
\nonumber\\ &&
\times \hspace{0.5mm} \sum_{T} \eta_{T}(r) \hspace{0.5mm} \Theta(u^{\rm C-max}_T - u) 
\int^{2 \mu^2_{\chi T} (u^2 + v^2_{\rm esc}) / m_T}_{m_{\chi} u^2 / 2} dE_R \hspace{0.5mm} \frac{d\sigma_T}{dE_R},
\label{eq:cap_rate}
\end{eqnarray}

\noindent with $v_{\rm esc}(r)$ the escape speed at position $r$ inside the Sun,  $w^2=u^2 + v^2_{\rm esc}(r)$ the WIMP speed at the target position, and $u^{\rm C-max}_T = v_{\rm esc}(r) \sqrt{\frac{4 m_{\chi} m_T}{(m_{\chi} - m_T)^2}}$ the maximum speed of a WIMP with mass $m_{\chi}$ for which capture through the interaction with a target $T$ is kinematically possible. 
For the number density profile $\eta_{T}(r)$ of the different target nuclei in the Sun we use the Standard Solar Model AGSS09ph~\cite{solar_model_Serenelli2009}. 
The equation above was derived in~\cite{Gould:1987} where it was pointed out that the calculation of the total capture rate includes a sum over all possible incoming WIMP directions, effectively averaging over the angular distribution. So, also the capture rate in Eq.~(\ref{eq:cap_rate}) depends on the same speed distribution $f(u)$ discussed in Section~\ref{sec:dd} for DD. 

\section{Halo--independent constraints on WIMP--nucleon interactions using the single stream method}
\label{sec:single_stream}

In the present Section we outline the single--stream method introduced in~\cite{Halo-independent_Ferrer2015}.
In the absence of enough information about the speed of WIMPs in the Solar system 
one can assume that the speed distribution is given by some generalised function $f(u)$ which is normalized as:
\begin{equation}
\int^{u_{\rm max}}_0 du \hspace{0.5mm} f (u) = 1 ,
\label{eq:f_u_norm}
\end{equation}
where $u$ is the asymptotic speed of a WIMP with respect to the Sun and $u_{\rm max}$ is its 
maximum possible value. Assuming that all WIMPs in the halo are gravitationally bound to the Galaxy, one can write 
$u_{\rm max} = u_{\rm esc} + v_\odot$ where $u_{\rm esc}$ is the escape speed (in the Galactic rest frame) 
form the Galaxy at the location of the Sun and $v_\odot$ is the speed of the Sun in the halo. 
In our study 
we take $u_{\rm esc} = 560$ $\rm km/s$~\cite{vesc_Smith2006, vesc_Piffl2013} 
and $v_\odot = 220$ $\rm km/s$~\cite{SHM_maxwell_Green2011} which 
imply $u_{\rm max} = 780$ $\rm km/s$. 
However, we will also discuss the effects of choosing a much larger 
$u_{\rm max}$ in Section \ref{sec:analysis}.

The capture rate of WIMPs in the Sun (i.e., Eq.~(\ref{eq:cap_rate})) can be written as, 
\begin{equation}
C_\odot = \int^{u_{\rm max}}_0 du \hspace{0.5mm} f(u) \hspace{0.5mm} H_C(u) ,
\label{eq:cap_rate_H_u}
\end{equation}
where, 
\begin{eqnarray}
H_C(u) &=& \left(\frac{\rho_\odot}{m_{\chi}}\right) \hspace{0.5mm} \frac{1}{u} \int^{R_\odot}_0 dr \hspace{0.5mm} 4 \pi r^2 \hspace{0.5mm} \hspace{0.5mm} \sum_{T} 
\eta_{T}(r) \hspace{0.5mm} 
\Theta(u^{\rm C-max}_T - u) \hspace{0.5mm} 
\int^{2 \mu^2_{\chi T} (u^2 + v_{\rm esc}^2(r)) / m_T}_{m_{\chi} u^2 / 2} dE_R \hspace{0.5mm} 
 \nonumber\\
&&   \frac{2 m_T}{4\pi}\frac{1}{2 J_T+1}\sum_{\tau\tau^{\prime}k}\left [ R_{0k}^{\tau\tau^{\prime}}+R_{1k}^{\tau\tau^{\prime}} (u^2+v^2_{\rm esc}(r)-u^2_{\rm min}(E_R))\right ] 
W_{T k}^{\tau\tau^{\prime}} ,
\label{eq:NT_H_u} 
\end{eqnarray} 
with $u_{\rm min}^2 = \frac{m_T E_R}{2 \mu_{\chi T}^2}$.

Similarly, for the expected number of nuclear recoil events in a direct detection experiment 
one can express Eq.~(\ref{eq:DD_event}) as,
\begin{equation}
R_{\rm DD} = \int^{u_{\rm max}}_0 du \hspace{0.5mm} f(u) \hspace{0.5mm} H_{\rm DD}(u) ,
\label{eq:DD_H_u}
\end{equation}
where,
\begin{eqnarray}
H_{\rm DD}(u) &=& M{\tau_{\rm exp}} \hspace{0.5mm} \left(\frac{\rho_\odot}{m_{\chi}}\right) \frac{1}{u}
\sum_{T} N_T \int^{2 \mu^2_{\chi T} u^2 / m_T}_{0} dE_R \hspace{0.6mm} \zeta_T \nonumber\\
&& \frac{2 m_T}{4\pi}\frac{1}{2 J_T+1}\sum_{\tau\tau^{\prime}k}\left [ R_{0k}^{\tau\tau^{\prime}}+R_{1k}^{\tau\tau^{\prime}} (u^2-u^2_{\rm min}(E_R))\right ] 
W_{T k}^{\tau\tau^{\prime}}.
\end{eqnarray}

Considering one effective coupling (say, $c_i$) at a time, the expected number of events in a given DD experiment or the expected WIMP capture rate in the Sun can be written as:
\begin{equation}
R(c^2_i) = \int^{u_{\rm max}}_0 du f(u) H(c^2_i, u) 
\leq R_{\rm max} ,
\label{eq:rate}
\end{equation}
with $H$ being either $H_{\rm DD}$ or $H_C$ and $R_{\rm max}$ the corresponding bound. 
Since the response function $H(c^2_i, u)$ 
is proportional to $c^2_i$, one can write $H(c^2_i, u) = c^2_i H(c_i = 1, u)$, so that given the experimental upper bound $R(c_i^2)\le R_{\rm max}$ one obtains the relation:
\begin{equation}
R(c^2_i) = \int^{u_{\rm max}}_0 du f(u) \frac{c^2_i}{{c^2_i}_{\rm max}(u)} H({c^2_i}_{\rm max}(u), u) = 
\int^{u_{\rm max}}_0 du f(u) \frac{c^2_i}{{c^2_i}_{\rm max}(u)} R_{\rm max} \leq R_{\rm max} ,
\end{equation}
\noindent with:
\begin{equation}
H({c^2_i}_{\rm max}(u), u) = {c^2_i}_{\rm max}(u) H(c_i=1, u) = R_{\rm max},
\label{eq:c_max_u}
\end{equation}
\noindent from which one obtains the following upper bound on the coupling $c_i$ :
\begin{equation}
c^2_i \leq \left[\int^{u_{\rm max}}_0 du \frac{f(u)}{{c^2_i}_{\rm max}(u)}\right]^{-1} .
\label{eq:c2_upper_bound}
\end{equation}


Considering one neutrino telescope (NT) and one direct detection (DD) bound, 
in order to derive a halo--independent constraint 
three situations may occur: 

\begin{itemize}

\item \underline{Type I:}
\begin{eqnarray}
{({c^{\rm NT}})^2}_{\rm max}(u) &\leq& c^2_* 
\hspace{18mm} {\rm for} \hspace{2mm} 0 \leq u \leq \tilde{u}
\\
{({c^{\rm DD}})^2}_{\rm max}(u) &\leq& c^2_* 
\hspace{18mm} {\rm for} \hspace{2mm} \tilde{u} \leq u \leq u_{\rm max}
\end{eqnarray}
with ${({c^{\rm NT}})^2}_{\rm max}(u)$ and ${({c^{\rm DD}})^2}_{\rm max}(u)$ 
corresponding to ${c^2_i}_{\rm max}(u)$ for the NT and the DD experiments, respectively,
and $\tilde{u}$ the speed where 
${{c^{\rm NT}}}_{\rm max}$ and ${{c^{\rm DD}}}_{\rm max}$ intersect at a value $c_*$, i.e., 
${{c^{\rm NT}}}_{\rm max}(\tilde{u}) = {{c^{\rm DD}}}_{\rm max}(\tilde{u}) = c_*$. 
In this case, from Eq.~(\ref{eq:c2_upper_bound}) one can write~\cite{Halo-independent_Ferrer2015}: 
\begin{eqnarray}
c^2 &\leq& c^2_* \left[\int^{\tilde{u}}_0 du f(u) \right]^{-1} = \frac{c^2_*}{\delta}, 
\nonumber\\ 
c^2 &\leq& c^2_* \left[\int^{u_{\rm max}}_{\tilde{u}} du f(u) \right]^{-1} = \frac{c^2_*}{1-\delta} .
\end{eqnarray}
The quantity $\delta$ is defined as $\delta = \int^{\tilde{u}}_0 du f(u)$ 
(or $1-\delta = \int^{u_{\rm max}}_{\tilde{u}} du f(u)$). 
A unique halo--independent upper-limit on $c^2$ is obtained from the combination of 
a neutrino telescope and a DD experiment if $\delta = 1/2$. The corresponding 
upper-limit on the coupling is then:
\begin{equation}
c^2 \leq 2 \hspace{0.5mm} c^2_* .
\label{eq:limit_I}
\end{equation}

\item \underline{Type II:} 
In some cases it may happen that:
\begin{equation}
{({c^{\rm DD}})^2}_{\rm max}(u) > c^2_* \hspace{18mm} {\rm at} 
\hspace{2mm} u = u_{\rm max} .
\label{eq:condition_II}
\end{equation}
Then, following the procedure
presented in \cite{Halo-independent_Ferrer2015} the two conditions must apply:
\begin{eqnarray}
c^2 &\leq& c^2_* \left[\int^{\tilde{u}}_0 du f(u) \right]^{-1} = \frac{c^2_*}{\delta}, 
\nonumber\\ 
c^2 &\leq& {({c^{\rm DD}})^2}_{\rm max}(u_{\rm max}) \left[\int^{u_{\rm max}}_{\tilde{u}} du f(u) \right]^{-1} = \frac{{({c^{\rm DD}})^2}_{\rm max}(u_{\rm max})}{1-\delta} .
\end{eqnarray}
Now the halo--independent bound on $c^2$ is obtained if $\delta = \frac{c^2_*}{c^2_* + {({c^{\rm DD}})^2}_{\rm max}(u_{\rm max})}$, 
which implies that the halo--independent upper-limit on the coupling is:
\begin{equation}
c^2 \leq {({c^{\rm DD}})^2}_{\rm max}(u_{\rm max}) + c^2_* .
\label{eq:limit_II}
\end{equation}
Note that the condition~(\ref{eq:condition_II}) implies that in this case the conservative upper bound on the effective coupling becomes
sensitive to the choice of $u_{\rm max}$.

\item \underline{Type III:} Another possible situation may arise when the constraint obtained from a neutrino telescope dominates over that 
coming from a direct detection experiment throughout the entire range of speed, i.e., ${{c^{\rm NT}}}_{\rm max}(u) < {{c^{\rm DD}}}_{\rm max}(u)$ 
for $u \in [0, u_{\rm max}]$ and 
$\tilde{u} \not\in [0, u_{\rm max}]$. 
In this case, the halo--independent upper-limit on 
$c^2$, which is allowed by both the neutrino telescope and the DD experiment is obtained as, 
\begin{equation}
c^2 \leq {({c^{\rm NT}})^2} \left(u_{\rm max}\right) . 
\label{eq:limit_III}
\end{equation}
This happens mainly at lower $m_{\chi}$ for which even for a large WIMP incoming speed the capture rate remains  high, while the direct detection rate is suppressed. This is the only case when a conservative halo--independent bound on the coupling can be obtained without combining capture and DD. 
Note that here, too, the ensuing conservative halo--independent bound depends 
on the choice of $u_{\rm max}$.

\end{itemize}

For a given DM-nucleon interaction, at each $m_{\chi}$ we calculate the 
halo--independent upper-limit on the interaction coupling following 
either Eq.~(\ref{eq:limit_I}), (\ref{eq:limit_II}) or (\ref{eq:limit_III}), when appropriate.
In the case of more than one DD bound and NT bound, as in our analysis, the procedure described above must be repeated by combining each 
DD bound with each NT bound, and taking the most constraining halo--independent limit on $c_i$.

As pointed out above, in the situation described by type II the conservative bound becomes 
sensitive to the value of $u_{\rm max}$.
This effect can become important for a finite experimental energetic bin and/or when
the expected rate is suppressed at high recoil energies by the nuclear form factor. 
However, in the analysis of Section~\ref{sec:analysis} we will see that, 
for the experimental sensitivities that we are using here, 
for all the effective couplings considered in the present study 
type II is never realized, and the conservative bound is never sensitive to $u_{\rm max}$, 
unless $u_{\rm max}\gtrsim$ 8000 km/s, a value that largely exceeds the expectations for the escape speed in our Galaxy. In this sense the bounds obtained in Section~\ref{sec:analysis} will be halo independent. 


\begin{figure*}[ht!]
\centering
\includegraphics[width=7.49cm,height=6cm]{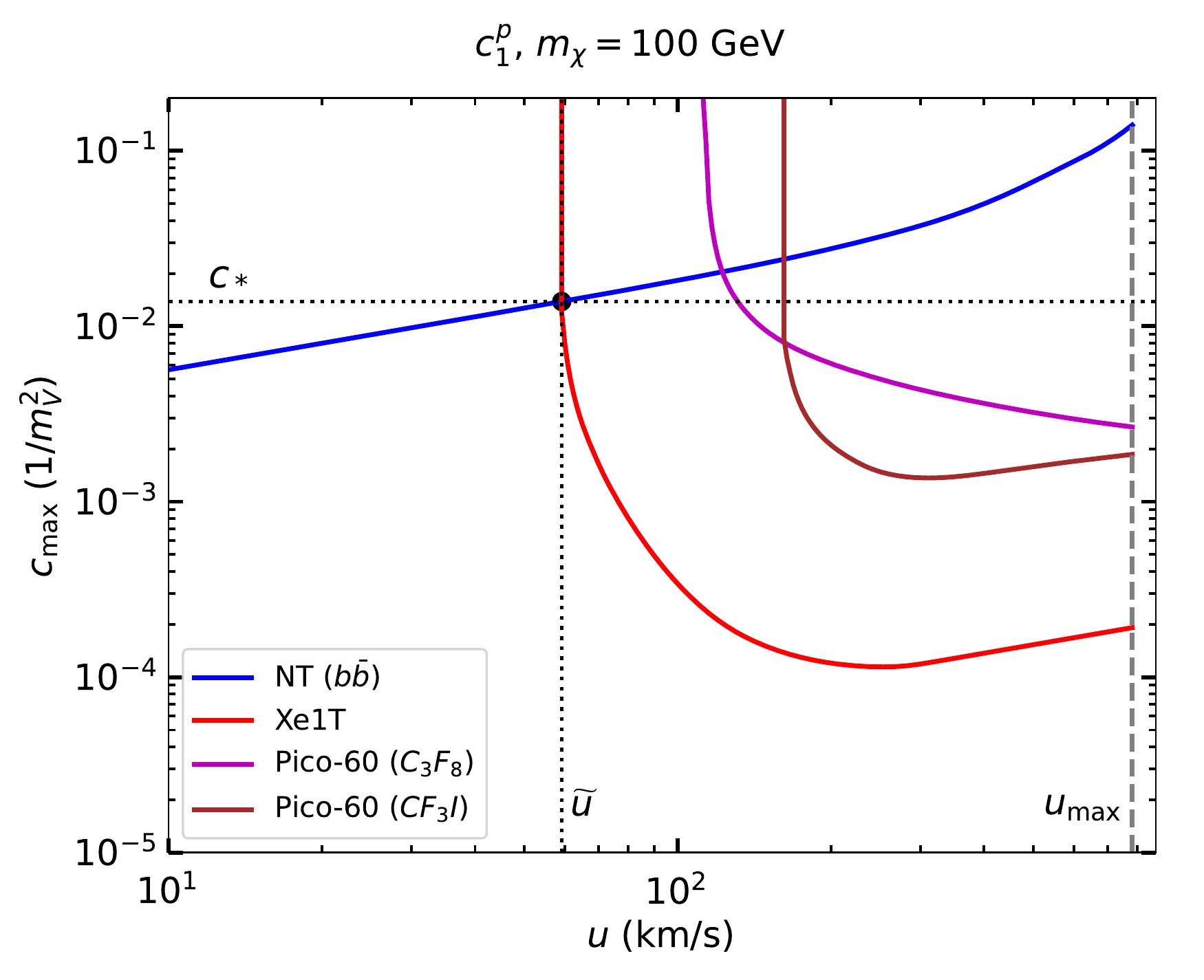}
\includegraphics[width=7.49cm,height=6cm]{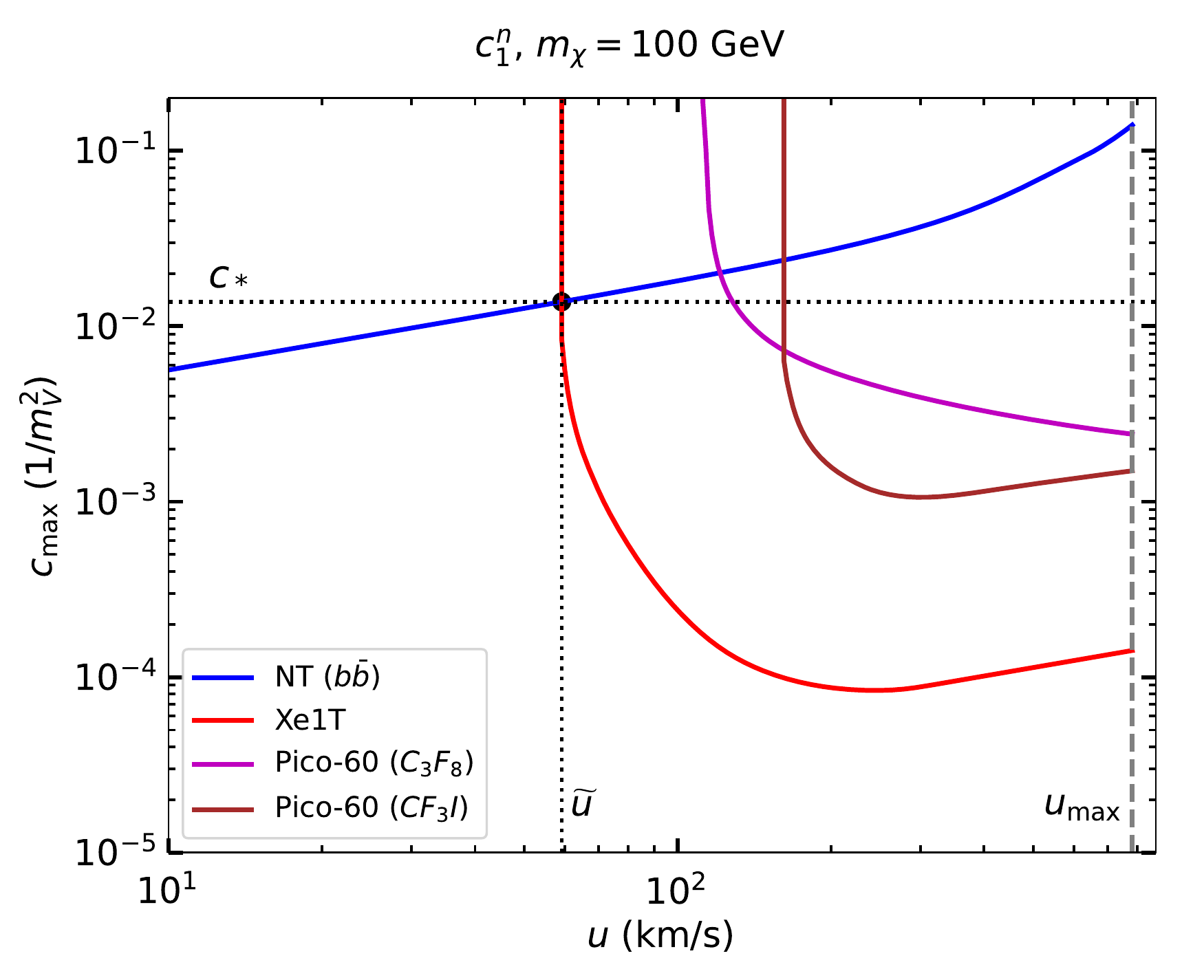}
\includegraphics[width=7.49cm,height=6cm]{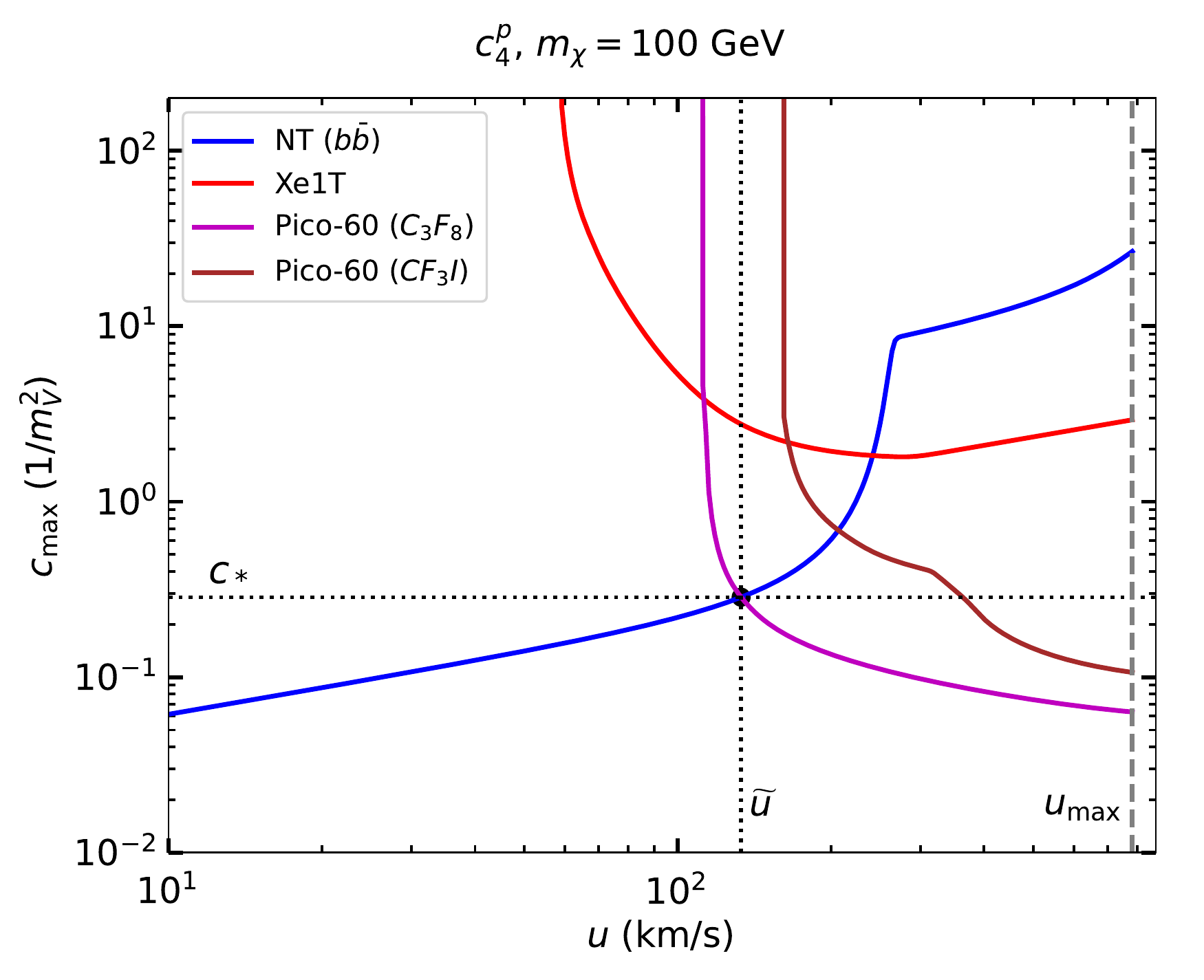}
\includegraphics[width=7.49cm,height=6cm]{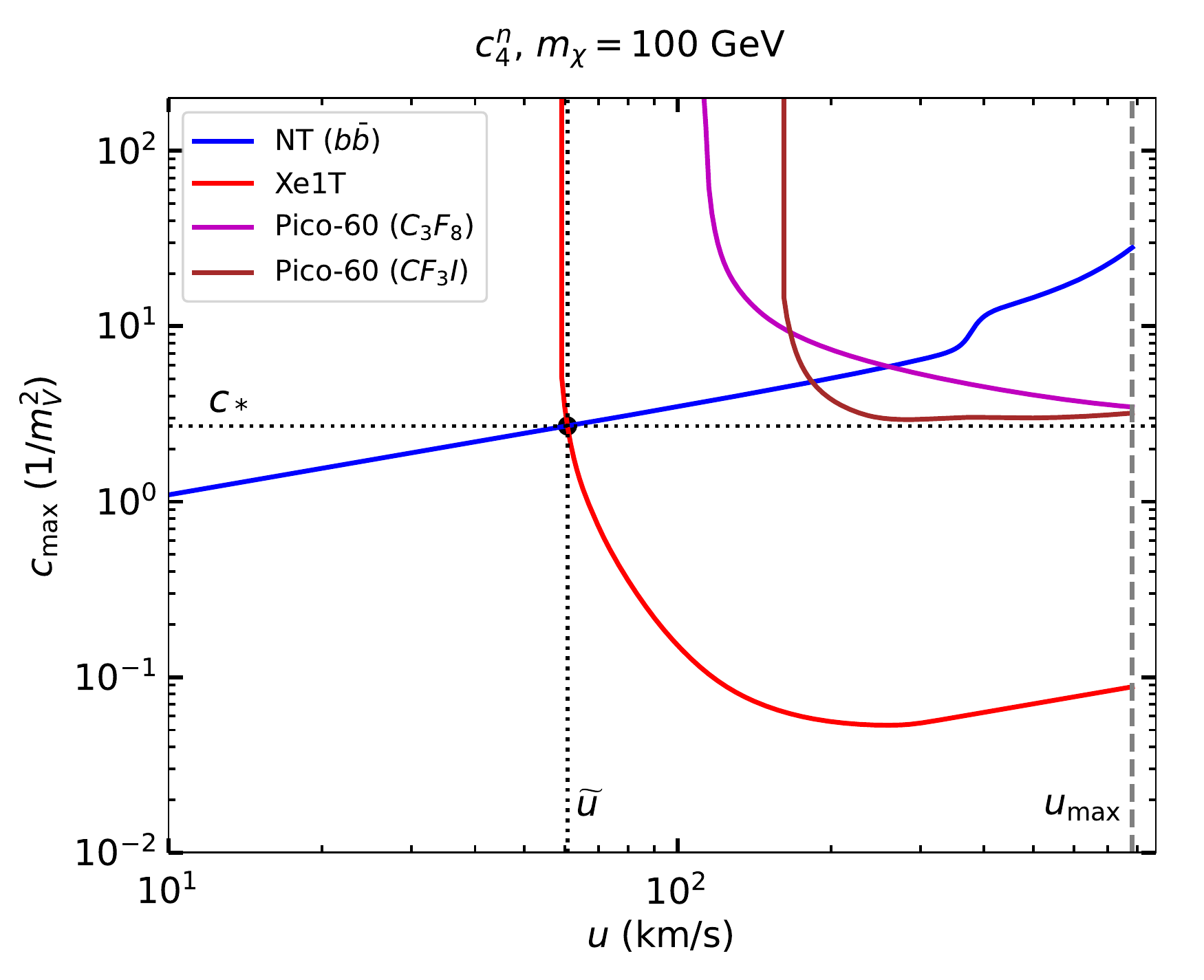}
\caption{Upper-limits ($c_{\rm max}$) on various non--relativistic DM-nucleon effective couplings estimated using different 
experiments are shown as a function of the WIMP stream speed $u$. 
At a given $u$ the quantity $c_{\rm max}$ for different experiments is 
calculated following Eq.~(\ref{eq:c_max_u}). 
For illustration, the limits are shown for a standard SI (top panels) or SD (bottom panels) interaction to the proton (left column) and the neutron (right column). In all cases the DM mass is chosen to be 
$m_{\chi} = 100$ GeV. To show the limits from neutrino telescopes (NTs) we use IceCube assuming $b\bar{b}$ as 
the dominant annihilation channel (blue curves), while for DD experiments we use 
XENON1T (red curves), PICO--60 ($C_3F_8$) (magenta curves) and PICO--60 ($CF_3I$) (brown curves). 
In each plot the maximum value of $c_{\rm max}$ allowed by all experiments simultaneously and the corresponding speed are indicated by $c_*$ and $\tilde{u}$, respectively. 
}
\label{fig:coupling_vel}
\end{figure*}

In Fig.~\ref{fig:coupling_vel} we show for a fixed WIMP mass $m_{\chi} = 100$ GeV 
the variation of the limit $c_{\rm max}(u)$ as a function of 
the WIMP speed $u$ for different experiments: 
neutrino telescope IceCube (blue curves, assuming $b\bar{b}$ as the dominant WIMP annihilation channel) 
and three DD experiments, i.e., XENON1T (red curves), PICO--60 ($C_3F_8$) (magenta curves) and 
PICO--60 ($CF_3I$) (brown curves). 
For illustration we assume four interactions, i.e., 
standard SI (${\cal O}_1$) and SD (${\cal O}_4$) interactions with both proton and neutron; the corresponding interaction couplings are indicated by $c^p_1$ (top left panel), $c^p_4$ (bottom left panel), 
$c^n_1$ (top right panel), $c^n_4$ (bottom right panel), respectively. 
In all the cases shown in Fig.~\ref{fig:coupling_vel} 
the halo--independent upper-limits on the couplings are given by Eq.~(\ref{eq:limit_I}). 
In each plot the point corresponding to $c_*$, representing the maximum value of $c_{\rm max}$ allowed by all experiments simultaneously, and the speed $\tilde{u}$ for which $c_*$ is obtained, 
is indicated by a black dot. In particular, for $c^p_1$, $c^n_1$ and $c^n_4$ the halo--independent limits 
are determined by the combination of IceCube and XENON1T, 
while for $c^p_4$ it is determined by IceCube and PICO--60 ($C_3F_8$).

\section{Analysis}
\label{sec:analysis}

\begin{figure*}[ht!]
\centering
\includegraphics[width=7.5cm,height=14cm]{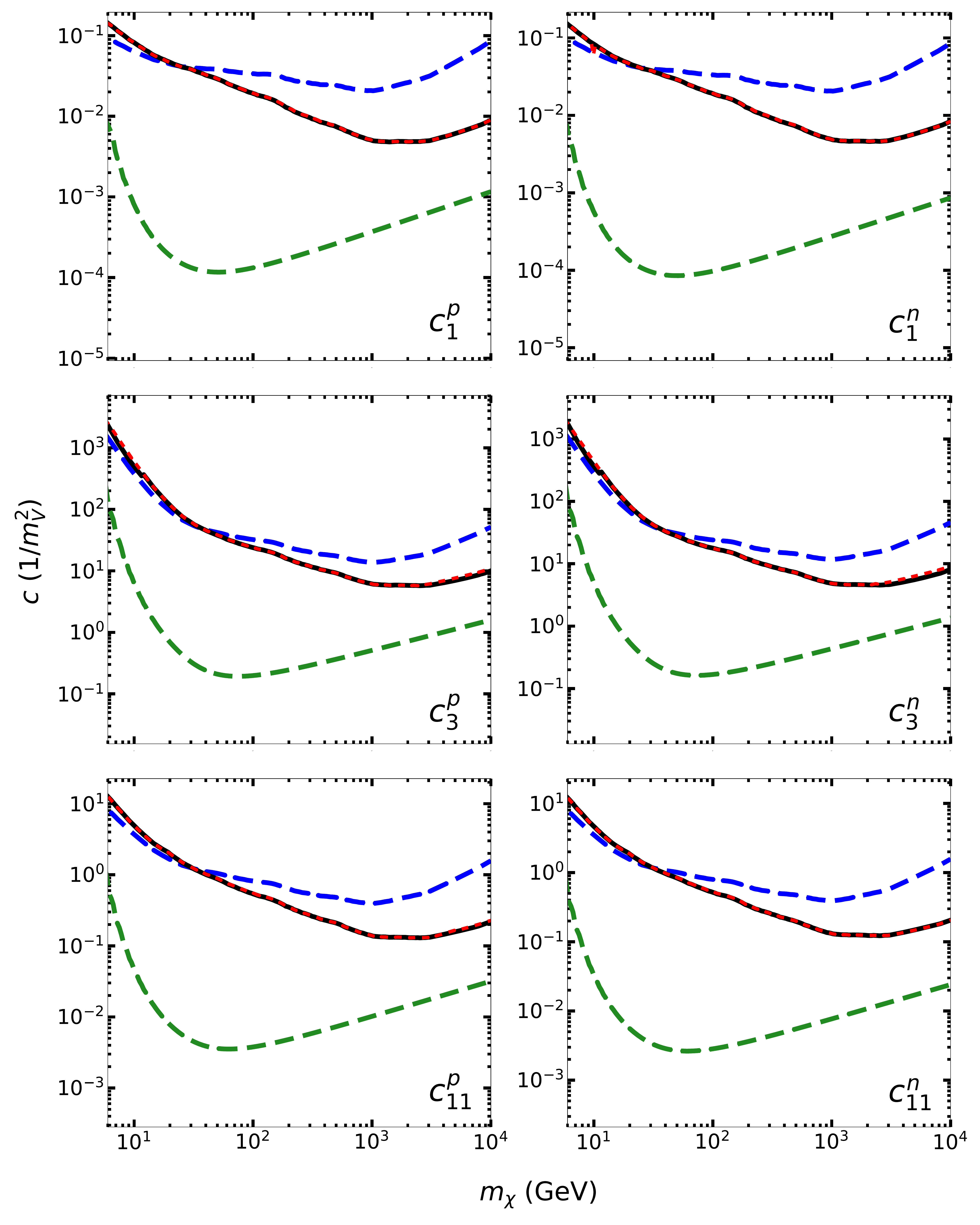}
\includegraphics[width=7.5cm,height=14cm]{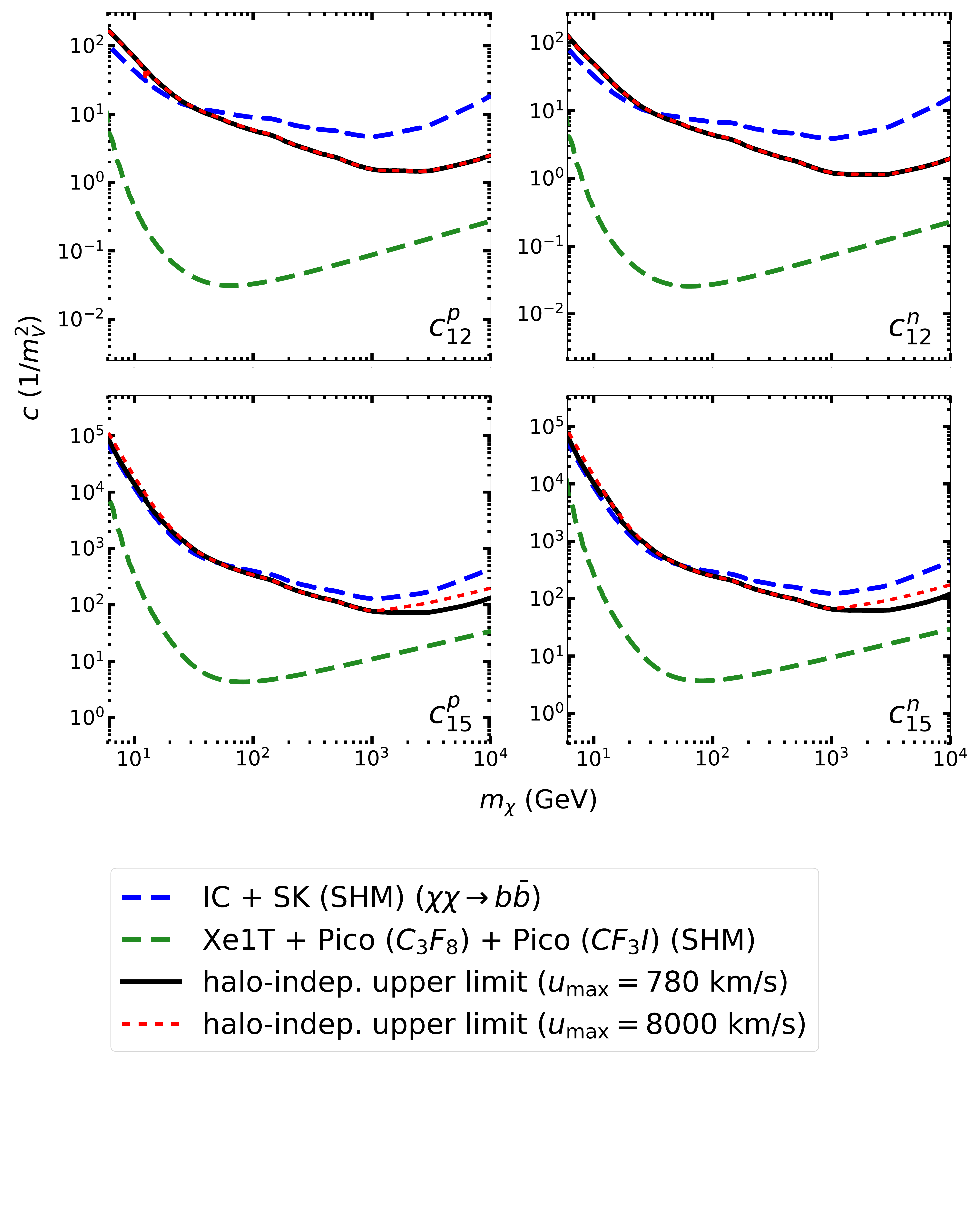}
\caption{Halo--independent upper-limits 
on different non--relativistic effective couplings $c_i$'s 
(for both proton and neutron) as a function of $m_{\chi}$, assuming 
$u_{\rm max} = 780$ km/s (the black solid lines)
as well as 8000 km/s (the red dashed lines). 
In this figure the interactions indicated in the text as ``Spin--Independent" 
(i.e., ${\cal O}_{1}$, ${\cal O}_{3}$, ${\cal O}_{11}$, ${\cal O}_{12}$ and ${\cal O}_{15}$) have been considered. 
The blue dashed lines show the combined upper-limits obtained from different 
neutrino telescope (NT) observations assuming the
Standard Halo Model and annihilations into $b\bar{b}$. 
The most stringent bound is from IceCube (IC)
for $m_{\chi} \gtrsim 100$ GeV and from Super-Kamiokande (SK) at lower WIMP masses. 
The green dashed lines represent the combined upper-limits (obtained assuming SHM) from 
three direct detection experiments: 
XENON1T, PICO--60 ($C_3F_8$) and PICO--60 ($CF_3I$).}
\label{fig:SI_coupling_mx}
\end{figure*}

\begin{figure*}[ht!]
\centering
\includegraphics[width=7.5cm,height=19.8cm]{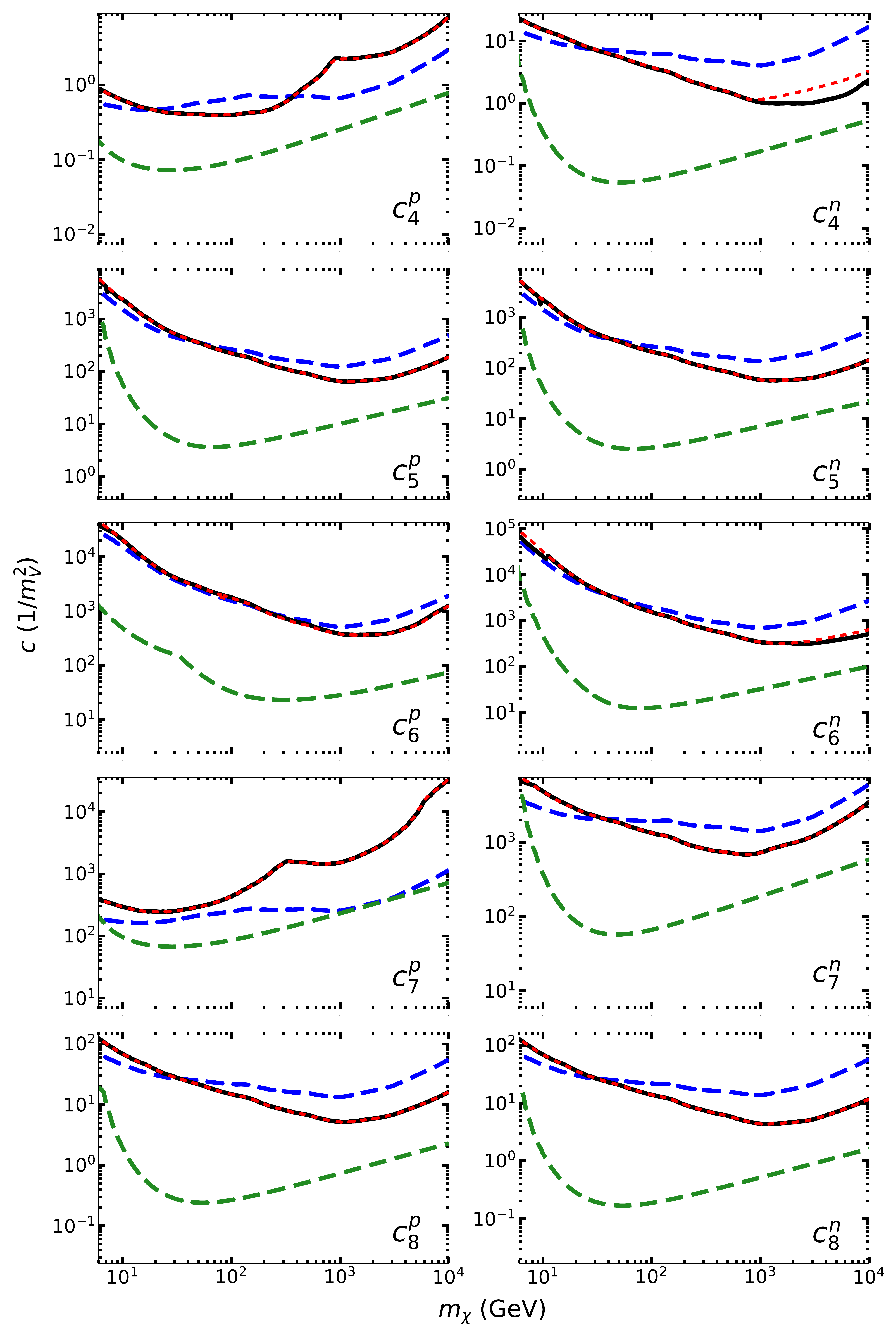}
\includegraphics[width=7.5cm,height=19.8cm]{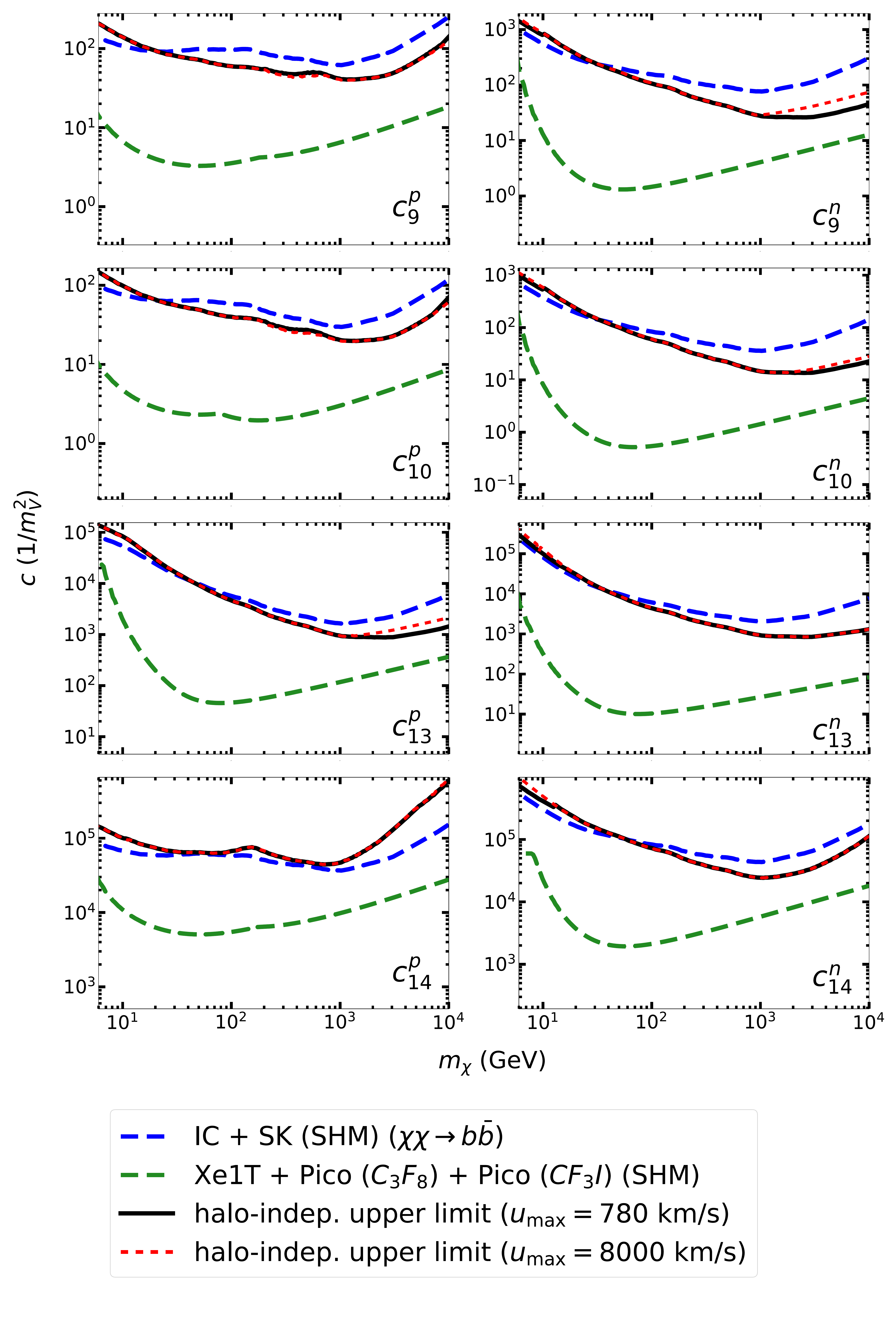}
\caption{The same as Fig.~\ref{fig:SI_coupling_mx}, but considering the interactions indicated in the text
as ``Spin–Dependent" (i.e. ${\cal O}_{4}$ -- ${\cal O}_{10}$, ${\cal O}_{13}$ and ${\cal O}_{14}$).}
\label{fig:SD_coupling_mx}
\end{figure*}

Neutrino flux predictions at Earth from WIMP annihilations in the Sun have been widely studied in the literature (see for instance~\cite{Blennow_2007}).
Experimental collaborations usually provide their bounds assuming that the WIMP annihilates predominantly to
$W^+W^-$, $\tau^+\tau^-$ or $b\bar{b}$~\cite{SuperK_2015, IceCube:2016, IceCube:2021, IceCube:2021_LE}. 
With the goal to obtain conservative bounds, in the following we will consider only annihilations to $b\bar{b}$, which, among them, is the annihilation channel that provides the smallest neutrino flux at detection\footnote{Even smaller signals are expected from WIMPs annihilating predominantly to muons or light quarks, which are stopped in the solar plasma before decaying and produce neutrinos in the MeV range~\cite{ritz_seckel_88} that are challenging to detect with the experimental threshold of present neutrino telescopes~\cite{rott_2012}.}.  

Further details about how we implement the neutrino and DD bounds are provided in the Appendix~\ref{app:experiments}.

The main results of our analysis are shown in Figs.~\ref{fig:SI_coupling_mx} and \ref{fig:SD_coupling_mx}. 
In these figures for each of the 14 operators for a WIMP of spin 1/2 listed in Table~\ref{tab:operators} the corresponding halo--independent conservative upper bounds on $c^p$ (WIMP--proton coupling) and $c^n$ 
(WIMP--neutron coupling) estimated using the methodology outlined in Section~\ref{sec:single_stream} 
are plotted as a function of the WIMP mass $m_\chi$ and are indicated by the black solid curves. 
In addition, the upper-limits on each coupling obtained from NT and DD experiments 
considering the SHM, i.e., a standard Maxwell speed distribution 
in the Galactic rest frame ($f_M(u)$) are also shown 
by blue and green dashed curves, respectively. 
For $f_M(u)$ we assume a speed dispersion $u_{\rm rms}=270$ km/s, a Galactic escape speed 
$u_{\rm esc}$ = 560 km/s 
and we boost it in the solar frame assuming $v_\odot$ = 220 km/s~\cite{SHM_maxwell_Green2011}. 
For the SHM case we plot at each WIMP mass the most constraining bounds from the NT and DD experiments.

In the two figures the effective operators are grouped in two main classes, according to the type of nuclear form factor that drives the corresponding interaction. In particular Fig.~\ref{fig:SI_coupling_mx} refers to the operators ${\cal O}_1$, ${\cal O}_3$, ${\cal O}_{11}$, ${\cal O}_{12}$ and ${\cal O}_{15}$. As shown in Table~\ref{table:eft_summary}, for such operators the WIMP--nucleus scattering process is driven either by $W^{\tau\tau^{\prime}}_{M}$ or $W^{\tau\tau^{\prime}}_{\Phi^{\prime\prime}}$. Such interactions are both enhanced for heavy targets. Specifically, $M$ corresponds to the standard spin--independent coupling proportional to the square of the
nuclear mass number; on the other hand, $\Phi^{\prime\prime}$ is non--vanishing for all nuclei and favors heavier elements with large nuclear shell model orbitals not fully occupied. Its scaling with the nuclear target is similar to the SI interaction, albeit the corresponding nuclear response functions are about two orders of magnitude smaller. We will refer to the class of operators shown in Fig.~\ref{fig:SI_coupling_mx} as ``spin--independent"--type interactions. 

The second class of operators, whose conservative bounds are shown in Fig.~\ref{fig:SD_coupling_mx}, corresponds to ${\cal O}_4$, ${\cal O}_5$, ${\cal O}_6$, ${\cal O}_7$, ${\cal O}_8$, ${\cal O}_9$, ${\cal O}_{10}$, ${\cal O}_{13}$ and ${\cal O}_{14}$. For all such operators the WIMP--nucleus interaction requires a non--vanishing nuclear spin. In particular ${\cal O}_4$, ${\cal O}_6$, ${\cal O}_7$, ${\cal O}_9$, ${\cal O}_{10}$ and ${\cal O}_{14}$ are driven by either the nuclear form factor $W^{\tau\tau^{\prime}}_{\Sigma^{\prime\prime}}$ or by $W^{\tau\tau^{\prime}}_{\Sigma^{\prime}}$, which directly couple the WIMP to the nuclear spin (the sum $W^{\tau\tau^{\prime}}_{\Sigma^{\prime\prime}}+W^{\tau\tau^{\prime}}_{\Sigma^{\prime}}$ corresponds to the standard spin--dependent form factor~\cite{menendez}). We will refer to this class of operators as ``spin--dependent"--type interactions. As we will see, their relevance is related to the fact that the Sun is mostly made up of targets with spin (in particular, $^{1}H$ targets).

As pointed out in Section~\ref{sec:single_stream}, in some cases the conservative bound can be sensitive to the particular choice of the value of $u_{\rm max}$ (see Eq.~(\ref{eq:limit_II})). This effect becomes important when $(c^{\rm DD})_{\rm max}(u_{\rm max})$ is very large, or, equivalently, when the response function ${H}_{\rm DD}$ is suppressed at large speeds. As explained below, indeed when $u$ is large enough $(c^{\rm DD})_{\rm max}(u)$ 
diverges linearly with $u$.  
Hence, a choice of $u_{\rm max}$ significantly larger than that adopted in our analysis, $u_{\rm max}$ = 780 km/s, 
may weaken our bounds. 
In order to estimate this effect 
we repeat the analysis of Section \ref{sec:single_stream} taking a value ten times larger, $u_{\rm max} = 8000$ km/s.
The resulting bounds for different couplings 
are shown in Figs.~\ref{fig:SI_coupling_mx} and \ref{fig:SD_coupling_mx} 
by the red dashed lines. 
By comparing these bounds with those obtained for $u_{\rm max} = 780$ km/s 
(shown with the solid black lines)
one can see that the effect of $u_{\rm max}$ in determining the conservative halo--independent bounds 
is rather mild, reaching at most a factor $\lesssim 2$ at large $m_\chi$ for some of the couplings ($c^p_{15}$, $c^n_{15}$, $c^n_4$, $c^n_9$ and $c^p_{13}$). Values of $u_{\rm max}$ largely exceeding 8000 km/s would eventually require to extend the treatment of the present analysis to the relativistic regime\footnote{The value of $u_{\rm max}$ = 8000 km/s is not realistic and we adopt it only to show that the halo--independent bound is almost insensitive to $u_{\rm max}$. In particular, for a given value of $u$ the single--stream method assumes that the full incoming WIMP flux has that same speed. However, if any, only a small fraction of the WIMPs with such a high speed 
can contribute to the DM local density in the neighbourhood of the Sun~\cite{Ibarra_Herrera}.}.

\begin{figure*}[ht!]
\centering
\includegraphics[width=7.49cm,height=6cm]{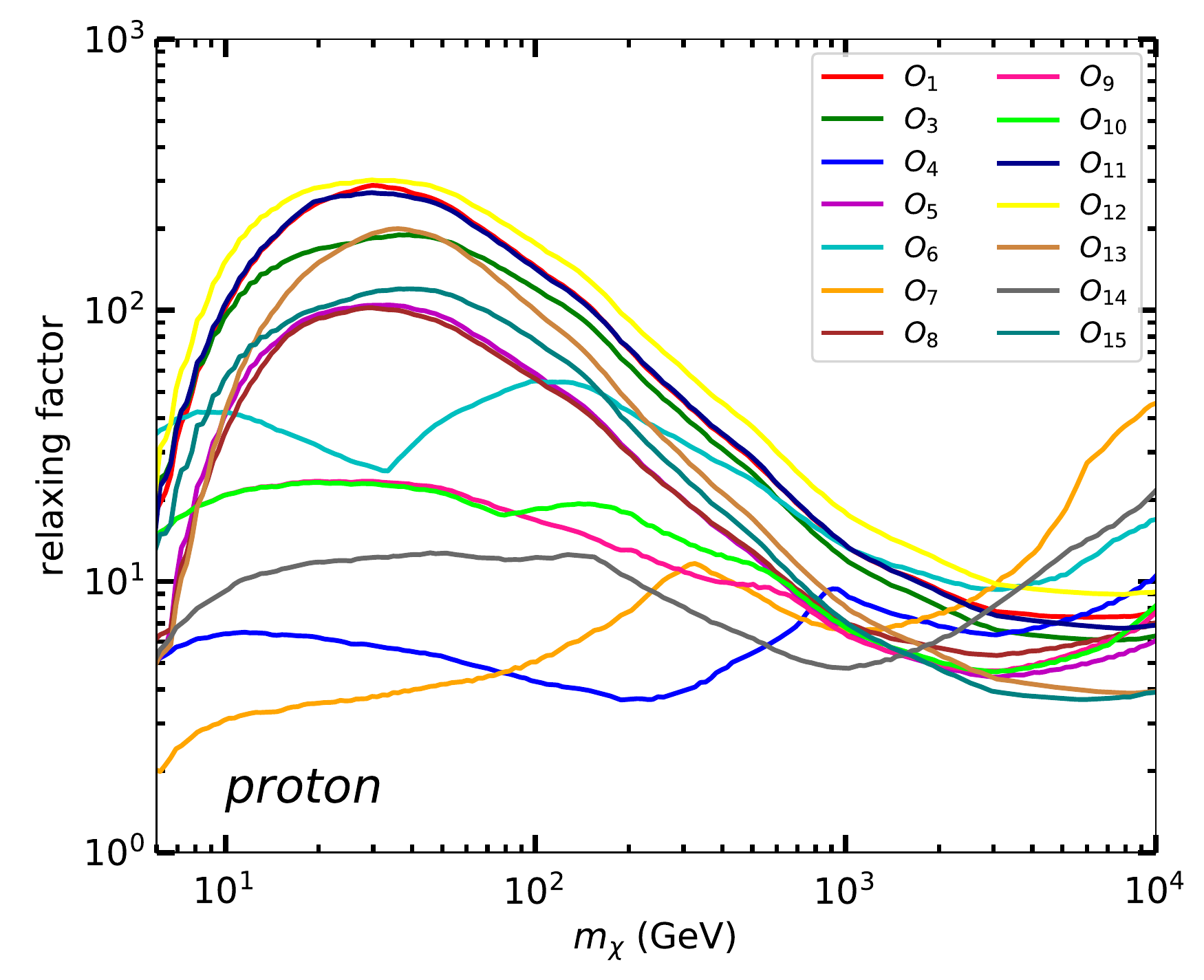}
\includegraphics[width=7.49cm,height=6cm]{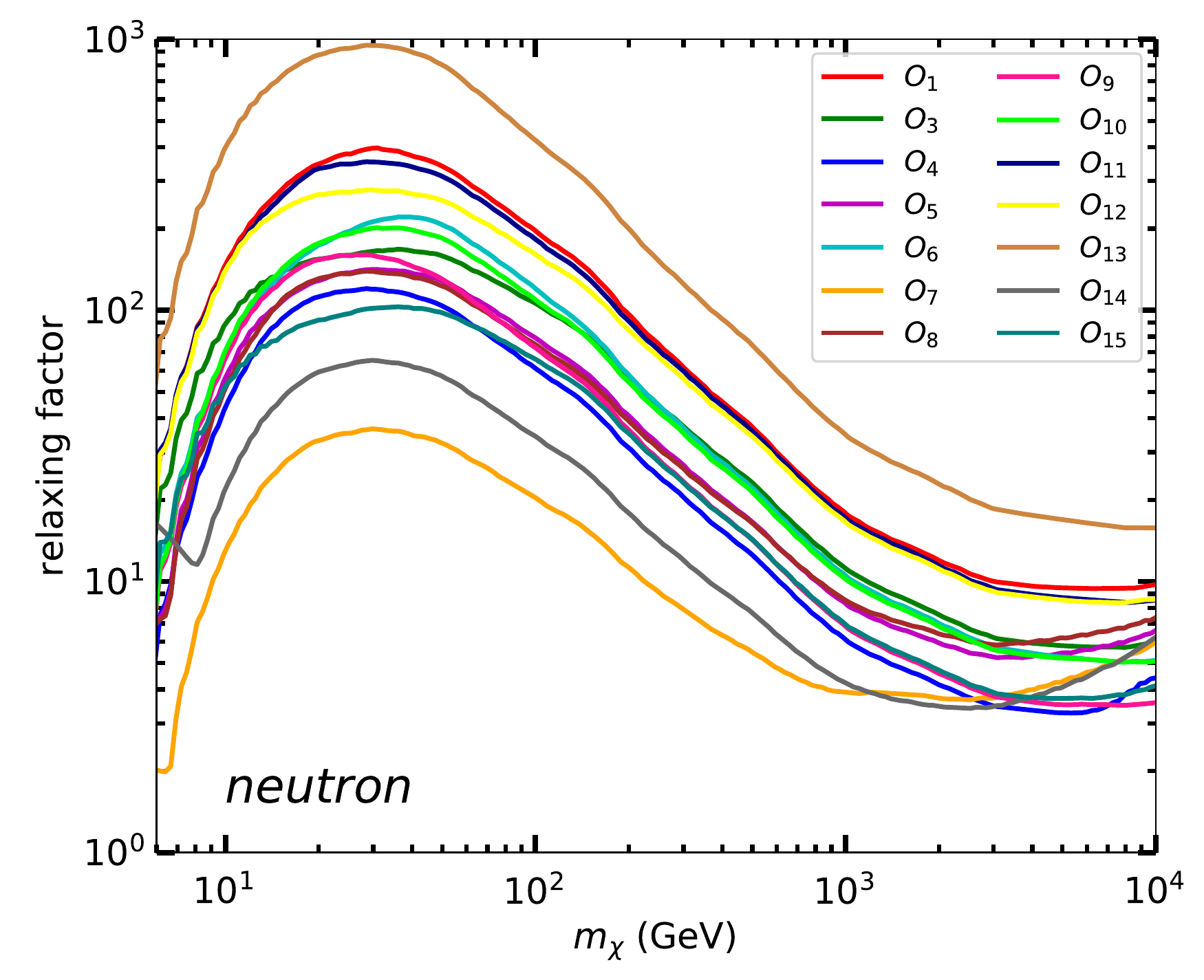}
\caption{The relaxing factor defined as $(c_i)_{\mbox{\small halo--indp.}} / (c_i)_{\mbox{\footnotesize SHM}}$ 
is plotted for each operator ${\cal O}_i$ as a function of $m_{\chi}$, 
considering couplings to the proton (left panel) and to the neutron (right panel). 
The quantity $(c_i)_{\mbox{\small halo--indp.}}$ 
is the halo--independent upper-limit on the coupling $c_i$ 
(the solid black line in Figs. \ref{fig:SI_coupling_mx} and \ref{fig:SD_coupling_mx}), 
while $(c_i)_{\mbox{\footnotesize SHM}}$ represents the strongest upper-limit on $c_i$ for a 
standard Maxwellian speed distribution.
}
\label{fig:relax_mx}
\end{figure*}

To compare how the halo--independent approach can affect the bounds on different effective operators, 
in Fig.~\ref{fig:relax_mx} we plot as a function of $m_\chi$ and for each of the couplings $c_i^p$ 
(left--hand plot) and $c_i^n$ (right--hand plot) a relaxing factor defined as the ratio between 
the following two quantities: 
the conservative exclusion plot obtained using the procedure outlined in Section~\ref{sec:single_stream} 
and plotted in Figs.~\ref{fig:SI_coupling_mx}, \ref{fig:SD_coupling_mx} (the black curve) 
and that obtained using for $f(u)$ a standard Maxwellian distribution $f_M(u)$
(the strongest one between the blue and the green dashed curves).  

In particular, the square of the relaxing factor $r_f$ shown in Fig.~\ref{fig:relax_mx} for a given coupling
is explicitly given by: 
\begin{equation}
r_f^2 = \frac{2 c^2_*}{(c^{\rm exp}_{\rm SHM})^2} = 2 c^2_* \int^{u_{\rm max}}_0 du 
\frac{f_{M}(u)}{({c^{\rm exp}})^2_{\rm max}(u)} = 
2c^2_* \hspace{1mm} {\left \langle \frac{1}{({c^{\rm exp}})^2_{\rm max}} \right \rangle}
\simeq 2 c^2_* \hspace{1mm} {\left \langle \frac{1}{({c^{\rm exp}})^2_{\rm max}} \right \rangle}_{\rm bulk} ,
\label{eq:rf}
\end{equation}
where ``exp" indicates the NT or DD experiment that 
provides the strongest upper-limit on the coupling at a given $m_{\chi}$ in the case of a standard Maxwellian speed distribution $f_{M}$. 
The brackets $\langle ...\rangle$ indicate an average weighted by the Maxwellian $f_{M}$ while  ${\langle ...\rangle}_{\rm bulk}$ indicates the dominant contribution to the average from the bulk of the WIMP speeds, defined as:
\begin{equation}
\int_{\rm bulk} du f_{M}(u) \simeq 0.8 .
\label{eq:bulk}
\end{equation}

From Fig.~\ref{fig:relax_mx} one can see that depending on the WIMP mass and on the effective operator the relaxation of the conservative bound compared to the standard Maxwellian case can either be as large as three orders of magnitude or as small as a factor of $\simeq$ 2.

The general features of the curves in Fig.~\ref{fig:relax_mx} can be understood in terms of the speed dependence of the quantities $(c^{\rm DD})^2_{\rm max}(u)$ and $(c^{\rm NT})^2_{\rm max}(u)$ that enter Eq.~(\ref{eq:c2_upper_bound}) and are proportional to $({H}_{\rm DD}(u))^{-1}$ and $({H}_C(u))^{-1}$, respectively. 

In particular $(c^{\rm DD})^2_{\rm max}(u)$ diverges at some speed threshold $u^{\rm DD}_{th}$ due to the experimental energy threshold (implemented in the detector response $\zeta_T$), and grows linearly with $u$ at large $u$. Specifically, this latter behaviour occurs for velocity--independent operators when the response function ${H}_{\rm DD}(u)$ corresponds to a finite energy bin $[E_{R,1},E_{R,2}]$ in the regime $u>u_{\rm min}(E_{R,2})$. However, for all the experiments that we include in our analysis the energy bin is not the main effect (in particular both PICO--60 ($C_3F_8$) and PICO--60 ($CF_3I$) are threshold detectors for which $E_{R,2}$ is not fixed). Instead, irrespective of whether or not an upper bound for the recoil energy is fixed in ${H}_{\rm DD}(u)$ and for both velocity--dependent and velocity--independent operators $(c^{\rm DD})^2_{\rm max}(u)$ always grows linearly with $u$ at large $u$. In fact, in the regime of large enough WIMP speeds the energy integral of Eq.~(\ref{eq:DD_H_u}) eventually stops depending on $u$ because the nuclear form factor $W_{Tk}^{\tau \tau^{\prime}}$ is suppressed for energies smaller than the $u$--dependent integration upper bound. Between the two asymptotic regimes described above $(c^{\rm DD})^2_{\rm max}(u)$ has a minimum where it is rather flat. 

As far as capture is concerned, $(c^{\rm NT})^2_{\rm max}(u)$ grows linearly with $u$ at low values of $u$ because in this regime the energy integral of Eq.~(\ref{eq:NT_H_u}) does not depend on $u$, due to the fact that in the lower bound $m_\chi u^2/2 \rightarrow$ 0 while the upper bound is almost insensitive on $u$ because the escape 
speed in the Sun $v_{\rm esc} \gg u$. This linear behaviour is eventually modified when 
$u \rightarrow u^{\rm C-max}$, where $(c^{\rm NT})^2_{\rm max}(u)$ diverges.

\begin{figure*}[ht!]
\centering
\includegraphics[width=4.9cm,height=4.3cm]{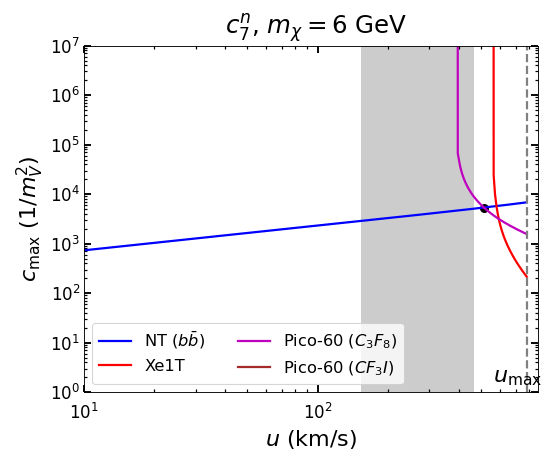}
\includegraphics[width=4.9cm,height=4.3cm]{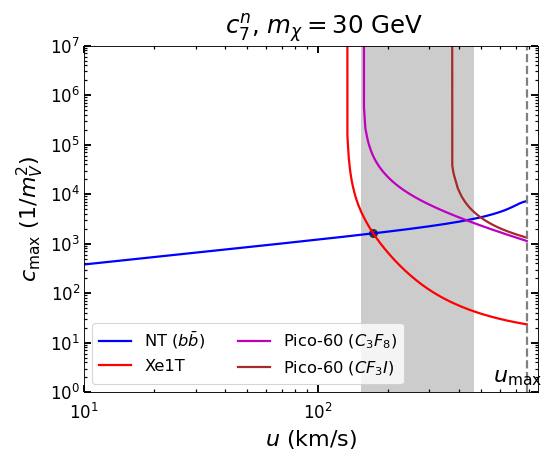}
\includegraphics[width=4.9cm,height=4.3cm]{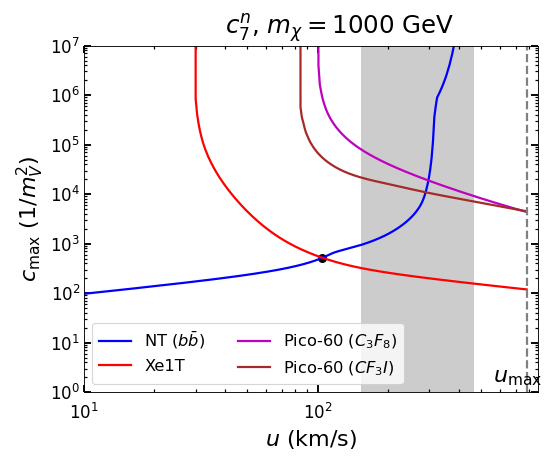}
\includegraphics[width=4.9cm,height=4.3cm]{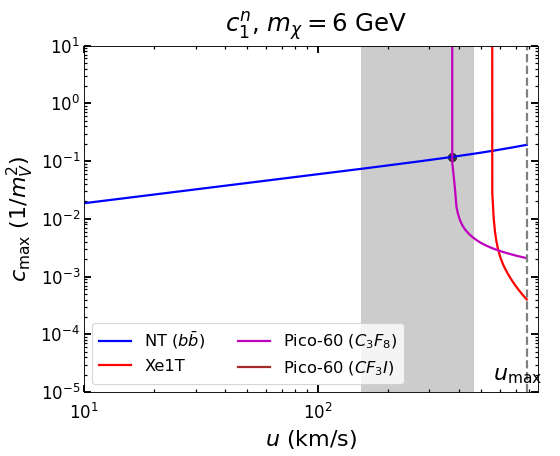}
\includegraphics[width=4.9cm,height=4.3cm]{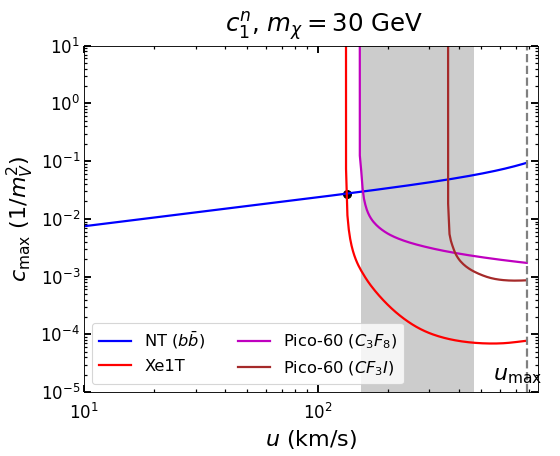}
\includegraphics[width=4.9cm,height=4.3cm]{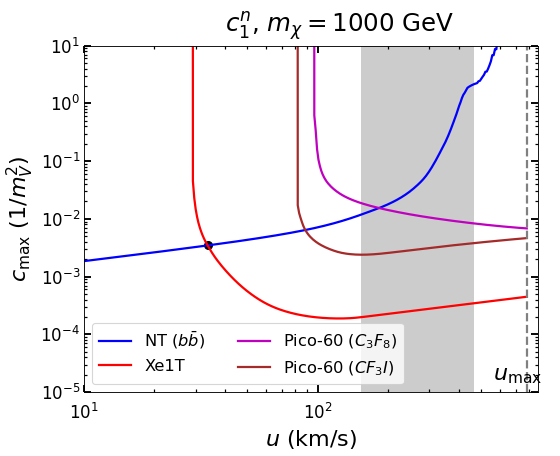}
\caption{$c_{\rm max}$ corresponding to different experiments are shown as a 
function of $u$ for three values 
of $m_{\chi}$, i.e., 6 GeV (left), 30 GeV (middle) and 1000 GeV (right) 
to explain the general behavior of the relaxing factor at different WIMP masses 
(shown in Fig.~\ref{fig:relax_mx}). Plots are shown for two representative WIMP-nucleon couplings: 
$c^n_7$ (top panel) and $c^n_1$ (bottom panel). 
The NT limits (blue curves) are obtained for 
Super-Kamiokande (for $m_{\chi} = $ 6 and 30 GeV) and IceCube (for $m_{\chi} = 1000$ GeV) assuming that $b\bar{b}$ 
is the dominant WIMP annihilation channel. 
The black dot in each plot indicates the corresponding ($\tilde{u}, c_*$).
The grey shaded region indicates the speed range which corresponds to the bulk of the 
Maxwellian speed distribution, i.e., the range where $\int du f(u) \simeq 0.8$ 
for a Maxwellian distribution. 
}
\label{fig:coupling_vel_2}
\end{figure*}

In particular, the behaviours described above for 
both $(c^{\rm DD})^2_{\rm max}(u)$ and $(c^{\rm NT})^2_{\rm max}(u)$ 
are shifted to higher speeds at small $m_\chi$, and to smaller speeds at large $m_\chi$. As a consequence, at small $m_\chi$ both $u^{\rm C-max}$ and $u^{\rm DD}_{th}$ are shifted to large values, so that $(c^{\rm NT})^2_{\rm max}(u)\lesssim(c^{\rm DD})^2_{\rm max}(u)$  with $(c^{\rm NT})^2_{\rm max}(u)$ rather flat with a linear behaviour up to $u_{\rm max}$, while $\tilde{u}$ is beyond the Maxwellian bulk region or close to its upper edge.
In this case $(c^{\rm NT})^2_{\rm max}(u)$ remains flat in a speed range that includes both the bulk of the Maxwellian and $\tilde{u}$ (see Fig.~\ref{fig:coupling_vel_2}), and as a consequence in Eq.~(\ref{eq:rf}) $c^2_*$ and $(\langle 1/(c^{\rm NT})^2_{\rm max} \rangle)^{-1}$
do not differ much, so that the relaxing factor is not large. 
This can be seen in Fig.~\ref{fig:relax_mx} at low values of $m_\chi$. We notice that, for values of 
$u_{\rm max}$ in agreement with present estimations of the escape speed, when $m_\chi$ is small enough $u^{\rm C-max} > u_{\rm max}$, i.e. capture probes alone the full range of expected WIMP speeds. As already pointed out, this is the only case when a conservative halo--independent bound on the coupling can be obtained without combining capture and DD. 

\begin{figure*}[ht!]
\centering
\includegraphics[width=4.9cm,height=4.3cm]{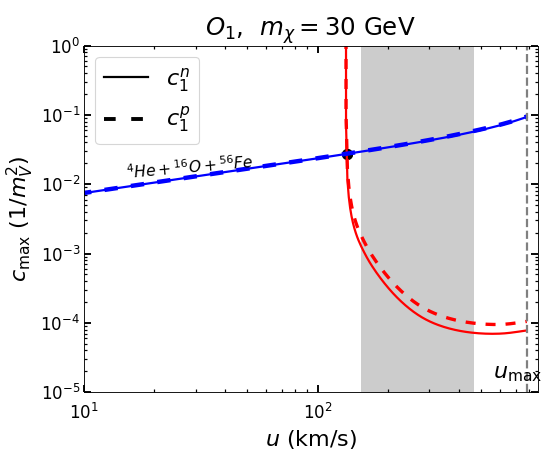}
\includegraphics[width=4.9cm,height=4.3cm]{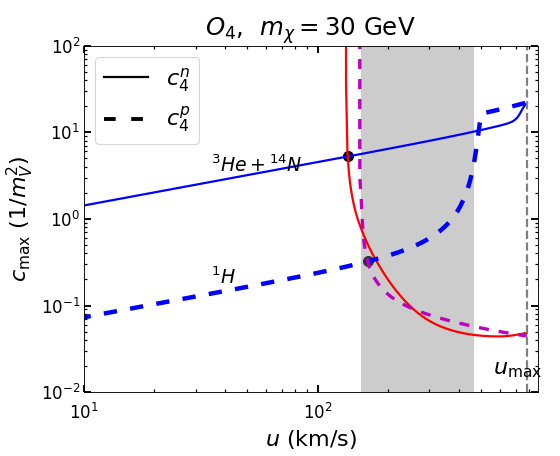}
\includegraphics[width=4.9cm,height=4.3cm]{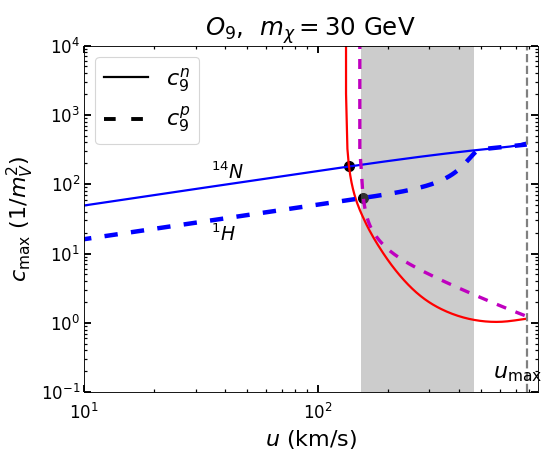}
\caption{$c_{\rm max}(u)$ is shown for ${\cal O}_1$ (left), 
${\cal O}_4$ (middle) and ${\cal O}_9$ (right) 
considering both WIMP--neutron (solid lines) and WIMP--proton (dashed lines) couplings 
to illustrate the effect of the presence of $^{1}H$ (or proton) 
in the Sun in reducing the relaxing factors 
of the WIMP--proton coupling of some SD operators for $m_{\chi} = 30$ GeV. 
For each coupling only the dominant DD experiment is shown, 
e.g., XENON1T (the red curve) for $c^n_1$, $c^p_1$, $c^n_4$ and $c^n_9$, 
and PICO--60 ($C_3F_8$) (the purple curves) for $c^p_4$ and $c^p_{9}$. 
The blue curves represent the limits corresponding to Super-Kamiokande 
assuming that $b\bar{b}$ is the dominant WIMP annihilation channel. 
The black dots indicate the intersections of NT and DD limits for the considered couplings.  
The grey shaded region indicates the speed range which corresponds to the bulk of the 
Maxwellian speed distribution, i.e., the range where $\int du f(u) \simeq 0.8$ 
for a Maxwellian distribution.}
\label{fig:coupling_vel_3}
\end{figure*}

In the opposite regime of large $m_\chi$ both $u^{\rm DD}_{th}$ and $u^{\rm C-max}$ are shifted to small values and in the Maxwellian case the bound is driven by DD. In this case $\tilde{u}$ is also driven to small values and below the bulk of the Maxwellian, where $(c^{\rm NT})^2_{\rm max}(u)$ is still linear or close to linear and intersects $(c^{\rm DD})^2_{\rm max}(u)$ 
just before the latter starts rising due to the $u^{\rm DD}_{th}$ threshold. In this case the range of speeds that includes $\tilde{u}$ and the bulk of the Maxwellian corresponds to the intermediate regime where $(c^{\rm DD})^2_{\rm max}(u)$ has a minimum and as a consequence is rather flat (see Fig.~\ref{fig:coupling_vel_2}). So also in this case in Eq.~(\ref{eq:rf}) the difference between $c^2_*$ and $(\langle 1/(c^{\rm DD})^2_{\rm max} \rangle)^{-1}$ is not large and the relaxing factor is moderate. This is observed in Fig.~\ref{fig:relax_mx} at large $m_\chi$, where for all the couplings (with the exception of  $c_6^p$, $c_7^p$ and $c_{14}^p$) the relaxing factor does not exceed one order of magnitude. 

Between the two asymptotic regimes of moderate relaxing factor at both small and large $m_\chi$ discussed above, 
in Fig.~\ref{fig:relax_mx} for most of the couplings the largest values of the relaxing factors are reached 
for 10 GeV $\lesssim m_\chi \lesssim$ 200 GeV, and specifically for $m_\chi\simeq$ 30 GeV. In this mass range the Maxwellian bound is driven by DD while $(c^{\rm NT})^2_{\rm max}(u)$ 
and $(c^{\rm DD})^2_{\rm max}(u)$ intersect where the latter has a steep dependence on $u$ because $u$ is close 
to $u^{\rm DD}_{th}$, with $\tilde{u}$ close to the lower edge of the bulk region (see Fig.~\ref{fig:coupling_vel_2}).
Due to these reasons in the range of speeds that includes $\tilde{u}$ 
and the bulk of the Maxwellian $(c^{\rm DD})^2_{\rm max}(u)$ changes significantly, so that
$(c^{\rm DD})^2_{\rm max}(u) \ll c^2_*$ for WIMP speeds in the bulk of the Maxwellian, and the 
relaxing factor is large. 
An exception to this pattern is represented by some of the WIMP--proton couplings, 
that in Fig.~\ref{fig:relax_mx} appear flattened in a peculiar way. This flattening corresponds to cases for 
which WIMP capture on $^1H$ dominates, since protons are by far the most abundant target inside the Sun and 
are instrumental in bringing $(c^{\rm NT})^2_{\rm max}(u)$ down at small $u$. As a consequence such effect is 
important for spin--dependent couplings, albeit reduced for those suppressed by momentum dependence (since $q$ is small 
due to the large mismatch between $m_\chi$ and the proton mass). 
This is shown in Fig~\ref{fig:coupling_vel_3}, 
where we consider three operators, ${\cal O}_{1}$ (standard SI), ${\cal O}_{4}$ (standard SD) and 
${\cal O}_{9}$ (spin--dependent with $q^2$ suppression, see Table~\ref{table:eft_summary}), 
and for each of them compare the $(c^{\rm NT})_{\rm max}(u)$ obtained for a WIMP--proton coupling with that obtained for a WIMP--neutron coupling. 
In the case of $c^p_1$ the quantity $(c^{\rm NT})_{\rm max}$, which is not driven by scattering events off protons, 
remains almost the same as that obtained for $c^n_1$, and hence 
the relaxing factor does not change much. 
On the other hand, for $c^p_4$, thanks to capture off protons, 
$(c^{\rm NT})_{\rm max}$ and as a consequence the relaxing factor get significantly reduced. 
The case of $c^p_{9}$ is somewhat in between due to momentum suppression. 
So in Fig.~\ref{fig:relax_mx}, for $m_\chi$ around 30 GeV, the relaxing factor 
is moderate for spin--dependent proton couplings, with the smallest values given by $c_4^p$ and $c_7^p$ (spin--dependent with no momentum suppression) 
followed by $c_9^p$, $c_{10}^p$ and $c_{14}^p$, which have a $q^2$ suppression, and by $c_6^p$, which has a $q^4$ suppression.

\begin{figure*}[ht!]
\centering
\includegraphics[width=7.49cm,height=6cm]{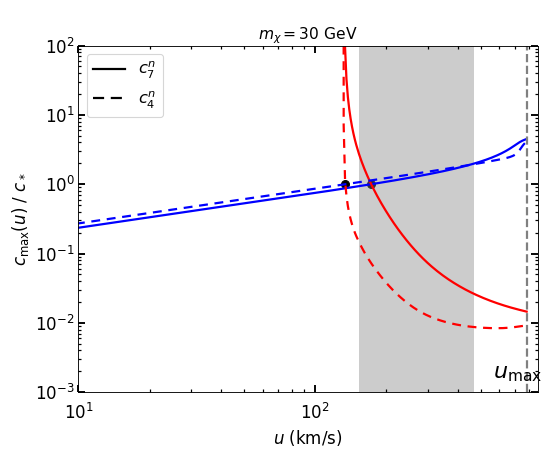}
\caption{$c_{\rm max}(u)$, normalised by $c_*$, is shown to explain 
the relatively small relaxing factor for the velocity--dependent operator ${\cal O}_{7}$ 
at $m_{\chi} = 30$ GeV
compared to that for the velocity--independent operator ${\cal O}_{4}$. 
Here only the WIMP-neutron interaction is considered and 
for both couplings ($c^n_7$ and $c^n_4$) only the dominant DD experiment, i.e., XENON1T (the red curves) 
is shown. 
The blue curves represent the limits corresponding to Super-Kamiokande assuming that 
$b\bar{b}$ is the dominant WIMP annihilation channel. 
The two black dots indicate the intersections of the NT and DD limits for the two considered couplings.
The grey shaded region indicates the speed range which corresponds to the bulk of the 
Maxwellian speed distribution, i.e., the range where $\int du f(u) \simeq 0.8$ 
for a Maxwellian distribution.}
\label{fig:coupling_vel_4}
\end{figure*}

We conclude our discussion with a few considerations about effective operators whose scattering amplitude only contains the velocity--dependent term (i.e. only $R_{1k}^{\tau\tau^{\prime}}$ in Eq.~(\ref{eq:r_decomposition}), see also Table~\ref{table:eft_summary}),  
namely ${\cal O}_7$ and ${\cal O}_{14}$. In particular our analysis shows that the relaxing factor of such operators somewhat differs compared to velocity--independent ones, albeit the effect is rather mild. In particular, in Fig.~\ref{fig:relax_mx} one can notice that at intermediate WIMP masses (around $m_\chi \simeq$ 30 GeV) the operators ${\cal O}_7$ and ${\cal O}_{14}$ have the smallest relaxing factors in the case of a WIMP--neutron coupling, and are among the smallest for a WIMP--proton one.  
One can notice that at this mass scale $c_*$ is determined by the intersection of a relatively flat
$(c^{\rm NT})_{\rm max}(u)$ and the rising part of $(c^{\rm DD})_{\rm max}(u)$ close to $u^{\rm DD}_{th}$, 
with the intersection speed $\tilde{u}$ close to the lower edge of the Maxwellian bulk. 
Due to the presence of the velocity square, 
the scattering cross section driven by $R_{1k}^{\tau\tau^{\prime}}$ (like in case of ${\cal O}_7$) has 
an extra suppression compared to that driven by $R_{0k}^{\tau\tau^{\prime}}$ (like in case of ${\cal O}_4$). 
However, for a given WIMP speed $u$ this suppression is more pronounced for the DD rate than 
for the capture rate because the latter is enhanced by the fact that its velocity dependence is mainly governed by the 
escape speed in the Sun $v_{\rm esc}$, 
which is typically much larger than $u$. This can be seen from Fig.~\ref{fig:coupling_vel_4} 
where we plot $c^{\rm NT}_{\rm max}(u)$ and $c^{\rm DD}_{\rm max}(u)$ 
normalised by the corresponding $c_*$ 
for operators ${\cal O}_7$ and ${\cal O}_{4}$ 
assuming a WIMP--neutron interaction.
With this normalisation the NT limits for both operators are almost at the 
same level in the full $u$ range, 
while the DD limit for ${\cal O}_7$ is comparatively weaker than that for ${\cal O}_4$. 
For this reason the value of $\tilde{u}$ is larger for ${\cal O}_7$ than for ${\cal O}_4$, a general feature 
for velocity--dependent operators compared to velocity--independent ones. 
As a consequence, for the former operator the difference between $c^2_*$ and 
$(\langle 1/(c^{\rm DD})^2_{\rm max} \rangle)^{-1}$ in the speed range that includes both $\tilde{u}$ and 
the Maxwellian bulk is comparatively smaller, leading to a lower relaxing factor.
This effect can be directly observed for a WIMP--neutron coupling, for which the smallest relaxing factors 
correspond to ${\cal O}_7$ and ${\cal O}_{14}$. 
However, in the case of a WIMP--proton coupling 
the neutrino signal from the Sun for ${\cal O}_{14}$ is driven by capture off protons, which is 
suppressed because of its explicit momentum dependence. This explains why in the left--hand plot of Fig.~\ref{fig:relax_mx} the relaxing factor for ${\cal O}_{14}$ at an intermediate WIMP mass is 
larger than that for ${\cal O}_{4}$, which does not contain any 
velocity--dependent term.

\section{Conclusions}
\label{sec:conclusions}

In the present work we have used the single--stream method introduced in Ref.~\cite{Halo-independent_Ferrer2015} to obtain halo--independent bounds on each of the 28 WIMP--proton and WIMP--neutron couplings of the effective non--relativistic Hamiltonian that drives the scattering process off nuclei of a WIMP of spin 1/2. The method only assumes that the speed distribution is normalized to one, but cannot be used when the process depends on more than one independent coupling. As a consequence, in our analysis we have considered a single non--vanishing coupling at a time. Since each of the non--relativistic operators of the effective Hamiltonian is a building block of the most general low-energy limit of any ultraviolet theory, such a  discussion is crucial for the interpretation of more general scenarios where  
several non--relativistic operators contribute to the signal.
 
In our analysis we have considered WIMP masses in the 6 GeV -- 10 TeV range, and have combined the updated null results from three direct detection experiments, XENON1T, PICO--60 ($C_3F_8$) and PICO--60 ($CF_3I$), along with those from two neutrino telescopes, Super-Kamiokande and IceCube. For the analysis of the latter we have assumed equilibrium between annihilation and capture, and taken the most conservative constraints obtained assuming $b\bar{b}$ as the dominant primary WIMP annihilation channel within the Sun.

Our main results are shown in Figs. \ref{fig:SI_coupling_mx} and \ref{fig:SD_coupling_mx}, 
where the halo--independent conservative bound for each coupling is plotted as a function of the WIMP mass, and compared to the corresponding combined limit from direct detection and capture in the Sun in the case of a Standard Halo Model for which the WIMP speed distribution is assumed to be a Maxwellian. In the same figures two values of the maximal WIMP speed $u_{\rm max}$ were considered, $u_{\rm max} = 780$ km/s (the usually expected escape speed of DM particles in the Galactic halo boosted to the solar system rest frame) and $u_{\rm max} = 8000$ km/s (a value ten times larger that largely exceeds the expected escape speed of DM particles) showing that the bounds in the latter case are at most relaxed by a factor of 2, 
and mainly in the large WIMP mass range. 
So the conservative bounds presented here can be considered as truly halo--independent, the only assumptions on the speed distribution and on the density of DM particles being that they are homogeneous at the the solar system scale.

In order to quantify the relaxation of the halo--independent bounds compared to those obtained by assuming the Standard Halo Model we have introduced a relaxing factor defined as the ratio between the halo--independent bounds derived with the single--stream method and the ones obtained for a Maxwellian speed distribution. We have found that, apart from some exceptional cases, the relaxing factor for most of the couplings shows a general pattern
(see Fig.~\ref{fig:relax_mx}). In particular, in the low and high $m_{\chi}$ regimes it is relatively moderate and in some cases it even becomes as small as a factor of $\simeq 2$. 
On the other hand in the intermediate mass range, 10 GeV $\lesssim m_\chi \lesssim$ 200 GeV, and for some of the couplings it can be as large as $\sim 10^3$. In such cases, given the lack of any direct evidence, besides numerical simulations, that the SHM correctly describes the halo of our galaxy, using the latter to derive bounds as commonly done in the literature appears to be an optimistic assumption. 

One main exception to the above mentioned general pattern of the relaxing factor is observed 
in the case of the WIMP--proton coupling of spin--dependent operators 
with no or a comparatively small ($q^2$) momentum suppression, namely ${\cal O}_{7}$, ${\cal O}_{4}$, ${\cal O}_{14}$, ${\cal O}_{9}$ and ${\cal O}_{10}$ (see Table~\ref{table:eft_summary}). In such cases the capture rate of WIMPs in the Sun is strongly enhanced because it is driven by WIMP scattering events off $^1H$, which is the most abundant element in the Sun, resulting in a more constraining bound that drives the relaxing factor down to the level of a few   
in the low and also in the intermediate $m_{\chi}$ regime. Within this class of operators and for the same momentum dependence the enhancement of solar capture compared to direct detection is largest for interactions that are driven by only the velocity--dependent term proportional to $w^2$ (${\cal O}_{7}$, ${\cal O}_{14}$), because of the highest speed of scattering WIMPs within the Sun, thanks to the large gravitational acceleration, compared to those close to the Earth's surface. In particular, indicating in parenthesis  the $q$ and $w$ dependences of the relevant cross sections,  the smallest relaxing factors between the halo--independent bound and that obtained using the SHM correspond in our analysis to the three couplings $c_4^p$ ($q^0$, $w^0$), $c_7^p$ ($q^0$, $w^2$) and $c_{14}^p$ ($q^2$, $w^2$), followed by  $c_9^p$ ($q^2$, $w^0$), $c_{10}^p$ ($q^2$, $w^0$) and $c_6^p$ ($q^4$, $w^0$), in the mass range 6 GeV $\lesssim m_\chi \lesssim$ 1 TeV.



\section*{Acknowledgements}
This research was supported by the National
Research Foundation of Korea(NRF) funded by the Ministry of Education
through the Center for Quantum Space Time (CQUeST) with grant number
2020R1A6A1A03047877 and by the Ministry of Science and ICT with grant
number 2021R1F1A1057119.

\appendix

\section{Implementations of experiments}
\label{app:experiments}

\subsection{XENON1T}
\label{app:xenon1t}

We reproduce relatively well the bound published in~\cite{xenon_2018} assuming an exposure of 278.8 days, a fiducial volume of 1.3 ton, and 7 WIMP candidate events in the range, 1.8 PE $ \le S_1 \le $ 62 PE, for which the efficiency is directly provided in terms of the nuclear recoil energy $E_R$ in Fig.~1 of~\cite{xenon_2018}, including the effects of quenching and energy resolution. This allows to obtain the expected rate by directly convoluting the differential rate of Eq.~(\ref{eq:dr_der}) with such efficiency, implemented as $\epsilon(E_R)$ in Eq.~(\ref{eq:zt}). 

\subsection{PICO--60 ($C_3F_8$)}
\label{exp:pico60_c3f8}

Bubble chambers are threshold experiments that detect a signal only 
above some value $E_{th}$ of the deposited
energy. In this case the expected number of
events is given by:

\begin{equation}
R=N_T MT\int_0^{\infty} P(E_R) \frac{dR}{dE_R} dE_R,
\label{eq:r_threshold}
\end{equation}

\noindent with $P(E_R)$ the nucleation probability.


One of the target materials used by PICO--60 is $C_3F_8$, for which we
used the complete exposure~\cite{pico60_2019} consisting in 1404 kg
day at threshold $E_{th}$=2.45 (with 3 observed candidate events and 1
event from the expected background, implying an upper bound of 6.42
events at 90\%C.L.~\cite{feldman_cousin}) and 1167 kg day at
threshold $E_{th}$=3.3 keV (with zero observed candidate events and
negligible expected background, implying a 90\% C.L. upper bound of
2.3 events).
For the $\epsilon(E_R)$ in Eq.~(\ref{eq:zt}) we have taken the nucleation probabilities of the two runs and each target element from Fig. 3 of \cite{pico60_2019}.

\subsection{PICO--60 ($CF_3I$)}
\label{exp:pico60cf3i}

PICO--60 can also employ a $CF_3I$ target. For the analysis of
Ref.\cite{pico60_2015} we adopt an energy threshold of 13.6 keV and an
exposure of 1335 kg days. The nucleation probabilities for each target
element are taken from Fig.~4 in~\cite{pico60_2015}.

\subsection{Neutrino Telescopes}
\label{exp:NT}

Neutrino telescopes put direct constraints on the neutrino flux from the 
annihilation of WIMPs captured in the Sun. Using the neutrino data taken from the direction of the Sun for a lifetime of 532 days the IceCube collaboration has provided a 90\% C.L. upper bound on the WIMP annihilation rate $\Gamma_\odot$ for different annihilation channels ($b\bar{b}$, $W^+W^-$ and $\tau^+\tau^-$) \cite{IceCube:2016}. In particular, for the $b\bar{b}$ channel $\Gamma_\odot\lesssim [7.4\times10^{24} \rm s^{-1}, 7.3\times10^{20} \rm s^{-1}]$ for $m_{\chi}$ in the range 35 GeV -- 10 TeV. 
The constraint from the Super-Kamiokande collaboration~\cite{SuperK_2015} is obtained using an exposure of 3903 days. The bound is expressed in terms of a 95\% C.L. upper-limit on the WIMP-nucleon cross section in the 6 -- 200 GeV WIMP mass range and corresponds to $\Gamma_\odot \lesssim [1.2\times10^{25} \rm s^{-1}, 1.2\times10^{23} \rm s^{-1}] $ for the $b\bar{b}$ primary annihilation channel.

%
The relation between  capture and annihilation in the Sun is given by $\Gamma_\odot$ = $(C_\odot / 2)$ $ {\rm tanh^2} (t_\odot / \tau_\odot)$, 
where $t_\odot$ and $\tau_\odot$ are the age of the Sun and the equilibrium time scale, respectively. In particular, if capture and annihilation are in equilibrium one has $\tau_\odot\ll t_\odot$ and $\Gamma_\odot$ = $(C_\odot / 2)$. In order to determine whether equilibrium is achieved in the Sun additional assumptions need to be made besides the non-relativistic effective theory. However, they are quite reasonable and common, and adopted in the ``standard" WIMP scenario: i) that the WIMP is a thermal relic providing the observed Dark Matter relic density with annihilation cross section times velocity at freeze--out of 
${\langle \sigma v \rangle} \simeq 3\times 10^{-26}$ cm$^{3}$ s$^{-1}$; ii) that the annihilation process is driven by an $s$--wave process, i.e. that the annihilation cross section is not velocity suppressed and the temperature dependence of $\langle\sigma v \rangle$ is negligible, so that the latter has the same value at WIMP freeze-out and inside the Sun. These pieces of information, together with the prediction of the capture rate from the non--relativistic effective theory, is sufficient to calculate the equilibration time $\tau_\odot$, which is given by~\cite{JUNGMAN1996}:

   \begin{equation}
    \tau_\odot=(C_\odot C_A)^{-1/2},
    \label{eq:tau_odot}
    \end{equation}
    with:

    \begin{equation}
        C_A=\frac{\langle\sigma v \rangle}{V_0} \left(\frac{m_\chi}{20 \mbox{GeV}} \right)^{3/2},
        \label{eq:ca}
     \end{equation}  
    and $V_0=(3 m_{PL}^2 T / (2 \rho \times 10 \mbox{GeV}))^{3/2}$, where $T$ and $\rho$ are the central temperature and the central density of the Sun, $T=1.4\times 10^7$ K, $\rho=150$ g$\cdot$ cm$^{-3}$.  This means that for a given WIMP mass $m_\chi$ and effective coupling $c_i$ the expected neutrino flux can be fully calculated including the effect of the equilibration time $\tau_\odot$, i.e. $\Gamma_\odot$ = $(C_\odot / 2)$ $ {\rm tanh^2} (t_\odot / \tau_\odot)$ = $\Gamma_\odot(m_\chi,c_i)$, and compared to the corresponding experimental upper bound, so that the assumption of equilibrium can be verified a posteriori and depends 
    on the experimental sensitivity. By direct substitution one can see that the IceCube and Super-Kamiokande upper bounds on $\Gamma_\odot$ correspond to values of the capture rate for which $t_\odot/\tau_\odot\gtrsim$ 150 (with $t_\odot\sim$ 4.7 Gy) so that $\tau_\odot\ll t_\odot$. For this reason we assumed equilibrium between capture and annihilation throughout our paper. 

Thanks to the lower threshold only Super-Kamiokande is sensitive to the low WIMP mass range, while IceCube is more constraining for $m_\chi$ heavier than a few tens of GeV. 
Our choice for the lower end of the $m_\chi$ interval (6 GeV) is due to the fact that we assume $b\bar{b}$ as the dominant primary final state of the annihilation process inside the Sun. Such value is anyway close to the evaporation lower limit $m_\chi \gtrsim$ 4 GeV~\cite{evaporation_Garani2017, evaporation_Busoni2017}. As far as the maximum value of $m_\chi$ (10 TeV) is concerned we adopted the same used in the IceCube experimental analysis. We notice that for such high WIMP masses neutrinos can be energetic enough to be partially absorbed in the Sun's medium, an effect included in the bound of IceCube.




\end{document}